\documentclass[11pt, onecolumn]{IEEEtran}
\usepackage{etex}
\usepackage{epsfig}
\usepackage{amsthm}
\usepackage[cmex10]{amsmath}
\usepackage{graphicx}
\usepackage{amssymb}
\usepackage{amsfonts}
\usepackage{psfrag,bm}
\usepackage{subfig}
\usepackage{multicol}
\usepackage[usenames,dvipsnames]{pstricks}
\usepackage{cancel}
\usepackage{tikz}
\usetikzlibrary{positioning}
\usepackage{epsfig}
\usepackage{pst-grad}
\usepackage{pst-plot}
\usepackage{dsfont}
\usepackage{authblk}
\usepackage{graphicx}
\usepackage{amsmath,amssymb,amsfonts,amsthm}
\usepackage{calc}
\usepackage{cases}
\usepackage{widetext}
\newtheorem{theorem}{Theorem}

\newtheorem{definition}{Definition}

\newtheorem{proposition}{Proposition}

\DeclareMathAlphabet{\mathpzc}{OT1}{pzc}{m}{it}
\renewcommand{\epsilon}{\varepsilon}

\newcommand{\txm}[1]{\textrm{\upshape #1}}

\newtheorem{rmk}{\sf Remark}

\newcommand{\Ca}{{\mathcal{C}}}

\makeatletter
\newcommand{\vast}{\bBigg@{4}}
\newcommand{\Vast}{\bBigg@{5}}
\makeatother

\def\b0{{\mathbf 0}}

\renewcommand{\baselinestretch}{2}
\begin{document}
\title{Multi-Phase Smart Relaying and Cooperative Jamming in Secure Cognitive Radio Networks}

\author{Pin-Hsun Lin, Fr\'{e}d\'{e}ric~Gabry, Ragnar Thobaben, Eduard A. Jorswieck,\\\vspace{-0.5cm} and Mikael Skoglund\\
\footnote{The material in this paper was presented in part at the IEEE Globecom 2014 Workshop - Trusted Communications with Physical Layer Security.\\
Pin-Hsun Lin and Eduard A. Jorswieck are with Communications Laboratory, Faculty of Electrical and Computer Engineering, Technische Universitat Dresden, Dresden, Germany (email: {Pin-Hsun.Lin, Eduard.Jorswieck}@tu-dresden.de).\\
Fr\'{e}d\'{e}ric~Gabry is with Mathematical and Algorithmic Sciences Lab, Huawei France Research Center, Paris, France (email:frederic.gabry@huawei.com).\\
Ragnar Thobaben and Mikael Skoglund are with School of Electrical Engineering and the ACCESS Linnaeus Centre, KTH Royal Institute of Technology, Stockholm, Sweden (email:ragnar.thobaben@ee.kth.se, skoglund@kth.se)\\
}}

\maketitle
\vspace{-3.5cm}
\begin{abstract}
In this paper we investigate cooperative secure communications in a
four-node cognitive radio network where the secondary receiver is
treated as a potential eavesdropper with respect to the primary
transmission. The secondary user is allowed to transmit his own
signals under the condition that the primary user's secrecy rate
and transmission scheme are intact. Under this setting we derive the
secondary user's achievable rates and the related constraints to
guarantee the primary user's weak secrecy rate, when
Gelfand-Pinsker coding is used at the secondary transmitter. In
addition, we propose a multi-phase transmission scheme to include 1)
the phases of the clean relaying with cooperative jamming and 2)
the latency to successfully decode the primary
message at the secondary transmitter. {A capacity upper bound for the secondary user is also derived.} Numerical results show that: {1)
the proposed scheme can outperform the traditional ones by properly
selecting the secondary user's parameters of different transmission
schemes according to the relative positions of the nodes; 2) the derived capacity upper bound is close to the secondary user's achievable rate within 0.3 bits/channel use, especially when the secondary transmitter/receiver is far/close enough to the primary receiver/transmitter, respectively.} Thereby,
a smart secondary transmitter is able to adapt its relaying and
cooperative jamming to guarantee primary secrecy rates and to
transmit its own data at the same time from relevant geometric
positions.
\end{abstract}

\section{Introduction}\label{sec:introduction}


Due to the broadcast nature of wireless networks, communications are
potentially subject to attacks, such as passive eavesdropping or active jamming. In contrast to the traditional cryptographic
approaches addressing these attacks \cite{cryp}, there
exists a promising direction towards
    achieving unconditional secure communications, namely {information-theoretic secrecy}. The information-theoretic
    secrecy approach, initiated by Shannon \cite{shaS} and
developed by Wyner \cite{Wyner75}, can exploit the
     randomness of the wireless channels to ensure the secrecy of the transmitted messages. As a
     performance measure for communication systems with secrecy constraints, a {secrecy rate} is defined
     as a rate at which the message can be transmitted reliably and securely between the legitimate nodes. However,
     similar to communication networks without secrecy constraints, the overall performance is limited by the relative channel
     qualities to guarantee secure communications. Many signal processing and multi-user techniques have therefore been proposed to overcome this limitation such
      as the use of multiple antennas \cite{ogg08,sha09,Khisti_MIMOME}.

Recently, there has been a substantial interest in the secrecy of
multi-user systems \cite{Liang2009}, with a particular emphasis on
potential cooperation between users to enhance the secrecy of
communications. Cooperation in communication networks is an emerging
technique to improve the reliability of wireless communication
systems, and it involves multiple parties assisting each other in
the transmission of messages, see e.g., \cite{tse04}. Assuming that
the cooperative node(s) can be trusted and that they aim at
increasing the secrecy of the original transmission in the presence
of a possible external eavesdropper, several cooperative strategies
have been proposed \cite{Gam07}, \cite{tek08}, \cite{poor10},
\cite{gabry14}. Many works have considered the impact of different variants of interference injection,
      such as noise-forwarding \cite{Gam07}, cooperative jamming (CJ) \cite{tek08}, or interference assisted secure
      communications
      \cite{intpoor}. While CJ with Gaussian noise has the advantage of simplicity, the non-decodability of the noisy signals are always hurting the legitimate receiver. Consequently, more elaborate CJ strategies have been recently proposed \cite{ulukus1} to mitigate this negative effect. The second type corresponds to the classical sense of cooperation,
where the cooperative nodes strengthen the main transmission by
using common relaying techniques such as decode-and-forward (DF),
amplify-and-forward \cite{poor10}, or compress-and-forward (CF)
\cite{kim11}. A comprehensive review of the main results for
multi-user networks with secrecy can be found in \cite{Liang2009}. {As one kind of cooperative communications schemes,} cognitive radio technology has been proposed by Mitola in
\cite{mit00} as an efficient way to enhance the spectrum efficiency which has considerable development over the last few decades. The state-of-the-art information theoretical analysis of cognitive radio systems can be found in \cite{rini1}, \cite{rini2}. The concept of cooperation for secrecy, and the corresponding
cooperative techniques can naturally be applied to the cognitive radio network.

{
In the present paper, we consider a four-node cognitive radio network
where the secondary receiver is treated as a potential eavesdropper
with respect to the primary transmission.  In exchange of
cooperation from the secondary user to improve his own secrecy rate,
the primary user allows the secondary user to share part of the
spectrum. Some important and related works are compared in the following. In \cite{CIC}, the secondary user wants to keep his message confidential to the primary
network. That is, the primary receiver is viewed as an eavesdropper from the secondary network
 perspective. Hence the CR transmitter
should make sure that the message is not leaked to the primary
receiver. In \cite{Somayeh}, the authors partially generalize the model of \cite{CIC} by additionally considering
the secrecy of the primary message and derive the rate
equivocation region of messages from both users. In \cite{farsani_SCRC_ISIT} the authors improve \cite{Somayeh} by rate splitting: part of the message is transmitted by the primary transmitter with a deterministic encoder, and the other part is transmitted by the cognitive
transmitter with a stochastic encoder. The main difference of our work to the aforementioned papers is that we additionally consider secure coexistence conditions to guarantee the primary user's secure transmission scheme is kept intact independently of the CR being active or not, which can simplify the design of the legacy system and also broaden the applicable usage scenarios. However, when considering these additional constraints, a deterministic encoder at the primary system \cite{CIC}, \cite{farsani_SCRC_ISIT} with rate splitting is incapable of guaranteeing the secrecy of the primary message. Furthermore, we investigate the inter-relation between channels observed by the primary transmitter when the cognitive transmitter is active or not to guarantee the secrecy constraint, which is a novel contribution compared to the cited papers.}

The main contributions\footnote{There are three main differences of the journal version to the previous conference version \cite{PH_GCWS14}: 1) We additionally derive the cognitive user's rate with dirty paper coding for discrete memoryless and AWGN channels, which treats the primary user's signal as a non-causally known side information.
2) We additionally derive a capacity upper bound of the considered model and compare the upper and lower bounds via numerical
results. The derived capacity upper bound is close to the achievable rate of the secondary user within 0.3 bits/channel use, especially when the secondary transmitter/receiver is far/close enough to the primary receiver/transmitter, respectively.
3) We additionally investigate a 4-phase scheme in which the relaying signal and jamming signal are transmitted in different phases, i.e., the third and fourth phases, respectively, to validate the choice of the proposed 3-phase scheme.} of this work are summarized as follows:
\begin{enumerate}
\item We analyze a cognitive radio network with the conditions that the secrecy rate as well as the transmission scheme of the primary network should be kept
intact for discrete memoryless channels. {One of the novel
applicable scenarios of the considered model is that the primary
system has no capability of being cognitive to the secondary users'
access and cannot adapt their transmission scheme accordingly, e.g.,
the commercial systems nowadays. Thus, besides the achievable rate
of the cognitive user\footnote{In this paper we use cognitive user and secondary user interchangeably.}, we derive the additional rate constraints to
guarantee that the primary user's weak secrecy is unchanged as well,
which requires different analysis compared to \cite{CIC}, \cite{Somayeh}, \cite{farsani_SCRC_ISIT}.}

\item We then propose a multi-phase transmission scheme, which considers the following additional phases. First, to accommodate the operations of practical systems, we take into  account
the first additional phase for listening to/decoding the primary's
signal at the secondary transmitter. Second, we introduce another additional phase as the {third
 one} to endow the cognitive system an extra degree
of freedom for utilizing different transmission schemes. For additive
white Gaussian noise (AWGN) channels, this degree of freedom
improves the performance by exploiting pure relaying and jamming but not simultaneously transmitting cognitive user's own signal. The relaying in this
interval is coined as \textit{clean relaying}.

{\item We derive a capacity upper bound (UB) for the secondary user under both discrete memoryless and AWGN channels to evaluate the performance of the achievable scheme.}


\item Finally, we illustrate our results through numerical examples based on a
geometrical setup, which highlights the impact of the node geometry
on the achievable rates and on the optimal power allocation and time
splitting of the secondary transmitter. Numerical results show that {1) the proposed 3-phase clean relaying scheme\footnote{In the following, we name the complete transmission scheme where the third phase uses the clean relay as the 3-phase clean relaying scheme.} indeed improves the cognitive user's rate; 2) the proposed achievable scheme is close to capacity when the secondary transmitter/receiver is far/close enough to the primary receiver/transmitter, respectively.}
\end{enumerate}
\vspace{0.5cm}
\textit{Notation:} In this paper, upper case normal alphabet denotes
random variables, lower and upper case bold alphabets denote vectors
and random vectors, respectively. The mutual information between two
random variables is denoted by $I(;)$. $E[.]$ denotes the
expectation; $H(.)$ and $h(.)$ denote the entropy and differential
entropy, respectively. $\sup$ and $\inf$ denote supremum and
infimum, respectively, $\Ca(x) \triangleq \log(1+x)$ and $a^+
\triangleq \max(0,a)$.

This paper is organized as follows. In Section~\ref{sec:systemModel}
we introduce our system model. In Section~\ref{sec:schemes} we
describe the transmission scheme and derive the achievable secrecy
rates for different signaling strategies. In Section ~\ref{sec:UB} we derive the capacity upper bound of the secondary user. Our theoretical results
are illustrated through numerical simulations in
Section~\ref{sec:numericalResults}. Finally,
Section~\ref{sec:conclusion} concludes this paper.

%
%
%

\section{System Model}\label{sec:systemModel}

\begin{figure*}
\begin{center}
\resizebox{0.99\textwidth}{!}{
\subfloat[][Complete model]{
\begin{tikzpicture}[scale=2.5]
\fill (0,0) node[circle,fill=black,minimum size=0.1pt, outer
sep=0pt,scale=0.5](t1) {} (0,0.2) node {T$_1$} (-0.15,-0.1) node
{$w_1$} (1.7,-0.3) node {$   =$}; \fill(0,-0.6) node[circle,fill=black,minimum
size=0.1pt,draw,outer sep=0pt,scale=0.5](t2) {} (-0.3,-0.8) node
{T$_2$} (0.1,-0.8) node {$w_2$ $(w_1)$}; \fill(1,0)
node[circle,fill=black,minimum size=0.1pt,draw,outer
sep=0pt,scale=0.5](u1) {}  (1,0.2) node {U$_1$} (1.3,0) node
{$w_1$}; \fill(1,-0.6) node[circle,fill=black,minimum
size=0.1pt,draw,outer sep=0pt,scale=0.5](u2) {} (1,-0.8) node
{U$_2$} (1.3,-0.6) node {$w_2$,$\xcancel{w_1}$};

\draw[->,line width=0.2mm] (t1)  -- node [above]{$1$} (u1) ;
\draw[->,line width=0.2mm] (t1)  -- node [left]{$c_{TT}$} (t2) ;
\draw[->,line width=0.2mm] (t2)  -- node [below]{$c_{22}$} (u2) ;
 \draw[->,line width=0.2mm] (t2)  -- node
[below=1.9pt,pos = 0.4]{$c_{21}$} (u1) ; \draw[->,line width=0.2mm]
(t1)  -- node [above=1.9pt, pos = 0.4]{$c_{12}$} (u2) ;
\end{tikzpicture}}

\subfloat[][The first Phase]{\label{First Phase}
\begin{tikzpicture}[scale=2.8]
%
\fill (0,0) node[circle,fill=black,minimum size=0.2pt,draw,outer
sep=0pt,scale=0.5](t1) {} (0,0.2) node {T$_1$} (-0.15,-0.2) node
{$\mathbf{x}_1^{(1)}$} (1.5,-0.3) node {$   +$};\fill(0,-0.6) node[circle,fill=black,minimum
size=0.2pt,draw,outer sep=0pt,scale=0.5](t2) {} (-0.1,-0.8) node
{T$_2$} (-0.15,-0.5) node {$\mathbf{y}_T^{(1)}$}; \fill(1,0)
node[circle,fill=black,minimum size=0.2pt,draw,outer
sep=0pt,scale=0.5](u1) {} (1,0.2) node {U$_1$} (1.3,0) node
{$\mathbf{y}_1^{(1)}$}; \fill(1,-0.6) node[circle,fill=black,minimum
size=0.2pt,draw,outer sep=0pt,scale=0.5](u2) {} (1,-0.8) node
{U$_2$} (1.3,-0.7) node {$\mathbf{y}_2^{(1)}$}; \draw[->,line
width=0.2mm] (t1)  -- (u1) ; \draw[->,line width=0.2mm] (t1)  --
(t2) ; \draw[->,line width=0.2mm] (t1)  --  node[above,pos=0.3]
{\textcolor{red}{(1)}} (u2) ;
\end{tikzpicture}
}
\subfloat[][The second Phase]{\label{Second Phase}
\begin{tikzpicture}[scale=2.8]
\fill (0,0) node[circle,fill=black,minimum size=0.1pt,draw,outer
sep=0pt,scale=0.5](t1) {} (0,0.2) node {T$_1$} (-0.15,-0.2) node
{$\mathbf{x}_1^{(2)}$} (1.5,-0.3) node {$   +$}; \fill(0,-0.6) node[circle,fill=black,minimum
size=0.2pt,draw,outer sep=0pt,scale=0.5](t2) {} (0,-0.8) node
{T$_2$} (-0.15,-0.5) node {$\mathbf{x}_2^{(2)}$}; \fill(1,0)
node[circle,fill=black,minimum size=0.2pt,draw,outer
sep=0pt,scale=0.5](u1) {} (1,0.2) node {U$_1$} (1.3,0) node
{$\mathbf{y}_1^{(2)}$}; \fill(1,-0.6) node[circle,fill=black,minimum
size=0.2pt,draw,outer sep=0pt,scale=0.5](u2) {} (1,-0.8) node
{U$_2$} (1.3,-0.7) node {$\mathbf{y}_2^{(2)}$}; \draw[->,thick,line
width=0.2mm] (t1)  --  (u1) ; \draw[->,thick,line width=0.2mm] (t2)
-- node[below]
{\textcolor{red}{(1)},\textcolor{green}{(2)},\textcolor{green}{(J)}}
(u2) ; \draw[->,line width=0.2mm] (t2)  --  node[below,pos=0.8] {
\hspace{4ex}
\textcolor{green}{(1)},\textcolor{red}{(2)},\textcolor{red}{(J)}}
(u1) ; \draw[->,line width=0.2mm] (t1)  --  node[above,pos=0.3]
{\textcolor{red}{(1)}} (u2) ;
\end{tikzpicture}
}
\subfloat[][The third Phase]{\label{Third Phase}
\begin{tikzpicture}[scale=2.8]
\fill (0,0)  node[circle,fill=black,minimum size=0.2pt,draw,outer
sep=0pt,scale=0.5](t1) {} (0,0.2) node {T$_1$} (-0.15,-0.2) node
{$\mathbf{x}_1^{(3)}$}; \fill(0,-0.6) node[circle,fill=black,minimum
size=0.2pt,draw,outer sep=0pt,scale=0.5](t2) {} (0,-0.8) node
{T$_2$} (-0.15,-0.5) node {$\mathbf{x}_2^{(3)}$}; \fill(1,0)
node[circle,fill=black,minimum size=0.2pt,draw,outer
sep=0pt,scale=0.5](u1) {} (1,0.2) node {U$_1$} (1.3,0) node
{$\mathbf{y}_1^{(3)}$}; \fill(1,-0.6) node[circle,fill=black,minimum
size=0.2pt,draw,outer sep=0pt,scale=0.5](u2) {} (1,-0.8) node
{U$_2$} (1.3,-0.7) node {$\mathbf{y}_2^{(3)}$}; \draw[->,line
width=0.2mm] (t1) --  (u1) ; \draw[->,line width=0.2mm] (t2)  --
node[below] {\textcolor{red}{(1)},\textcolor{green}{(J)}} (u2) ;
\draw[->,line width=0.2mm] (t2)  -- node[below,pos=0.8] {
\hspace{2ex} \textcolor{green}{(1)},\textcolor{red}{(J)}} (u1) ;
\draw[->,line width=0.2mm] (t1)  -- node[above,pos=0.3]
{\textcolor{red}{(1)}} (u2) ;
\end{tikzpicture}
}}

\caption{Multi-phase transmission scheme: (J) refers to jamming,
($1$) to $w_1$ and ($2$) to $w_2$. Green indicates a positive effect
(e.g. jamming, relaying) while red indicates a negative effect (e.g.
interference, eavesdropping).} \label{fig:2}
\end{center}
\end{figure*}
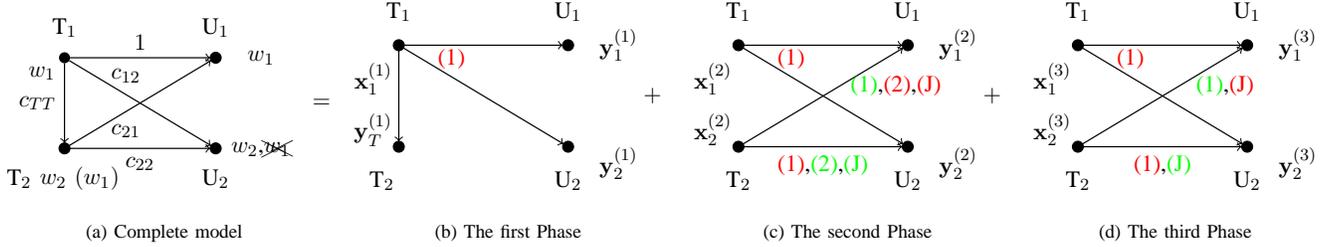
We investigate the cognitive radio channel with confidential message
(CRC-CM) at the primary user described in Fig. \ref{fig:2}, which
consists of the following single antenna half-duplex nodes: the
primary/secondary (cognitive) transmitters T$_{1}$/T$_{2}$ with
finite channel input alphabets $\mathcal{X}_1$ and $\mathcal{X}_2$,
respectively, and the primary/secondary receivers U$_{1}$/U$_{2}$
\footnote{Please note that we use the italic alphabets $U_1$ and
$U_2$ to denote the auxiliary random variables in Appendix.} with
finite channel output alphabets $\mathcal{Y}_1$ and $\mathcal{Y}_2$,
respectively. T$_{1}$ wishes to transmit the secret message $w_{1}$
to U$_{1}$, which should be kept secret from U$_{2}$. Meanwhile,
T$_{2}$ wants to transmit the message $w_2$ (without secrecy
constraints) to U$_2$. {In this work we consider the following two
requirements coined as the \textit{secure coexistence conditions}
inherited from \cite{Jovicic_CR}, which considers the coexistence
conditions without the secrecy constraint.

\begin{definition} \label{Def_CR}\it
The secure coexistence conditions require\\
(i) the transmission of T$_2$ does not degrade the primary user's secrecy
rate $R_{S1}$, and\\
(ii) the encoder and decoder at T$_1$ and U$_1$, respectively, are
left intact whether T$_2$ transmits or not.
\end{definition}

The reasons to consider the secure coexistence conditions are
twofold. First, to utilize the time-frequency slot in the overlay
sense, cognitive radio systems are obligated not to interfere the
primary systems, which is common in cognitive radio systems design.
Second, with the condition (ii), cognitive radios are backward
compatible to the legacy systems, which cannot sense and adapt to
the environment agilely. This condition makes the cognitive radio
capable of operating in broader scenarios. Note that the above
conditions are the main differences between our model and those in
\cite{CIC} \cite{Somayeh} \cite{farsani_SCRC_ISIT}. Furthermore, to attain the result, the
derivations to accommodate the secure coexistence conditions require different analysis. }

{One of the possible practical scenarios of the considered model is
that, the primary users belong to a licensed system, who sells
rights of the spectrum usage to a femtocell system. Here we can let the
secondary transmitter and receiver be the femtocell base station and
users, respectively. However, the femtocell operator may not be able
to guarantee that the femtocell users are malicious or not. Thus, to
provide a secrecy transmission to the primary users, not only the
primary base station needs to use the wiretap coding, but also the
femtocell base station needs to help to maintain that secrecy
transmission for the primary system. These considerations are
included into the secure coexistence conditions and will be
discussed in detail later.}

Denote $Y_1'/Y_2'$ and $Y_1/Y_2$ as the received
signals at U$_1$/U$_2$ when T$_2$ transmits or not,
respectively. A discrete memoryless CRC-CM is described as
\begin{align}\label{EQ_DMC1}
P(\bm y_1',\bm y_2'|\bm x_1, \bm x_2)=\overset{n}{\underset{k=1}\Pi}
P(  y_{1,k}',  y_{2,k}'|  x_{1,k},   x_{2,k}),
\end{align}
where the subscript $k$ denotes the $k$-th symbol, $ \bm x_1\in
\mathcal{X}_1^n$ and $ \bm x_2 \in \mathcal{X}_2^n$ are channel
inputs from T$_1$ and T$_2$, respectively; $ \bm y_1'\in
\mathcal{Y}_1^n$ and $\bm y_2'\in \mathcal{Y}_2^n$ are channel
outputs at U$_1$ and U$_2$, respectively, when T$_2$ transmits. The primary transmitter T$_1$ intends to send a message $W_1\in\mathcal{W}_1=\{1,\cdots,2^{nR_{S1}}\}$ to U$_1$ in $n$ channel uses. A $(2^{nR_{S1}},n)$ primary user's code is a choice of
encoding and decoding rules
\begin{align}
E_1^n:\,\{1,\cdots,2^{n R_{S1}}\}\rightarrow \mathcal{X}_1^n,\mbox{ and }
D_1^n:\, \mathcal{Y}_1^n \rightarrow \{1,\cdots,2^{n R_{S1}}\}.\notag
\end{align}
The weak secrecy rate $R_{S1}$ is achievable for the primary user for the wiretap channel if there exists a sequence of $(2^{nR_{S1}},n)$ codes such that the following two constraints are satisfied
\begin{align}
&\underset{n\rightarrow \infty}\lim
\frac{1}{2^{nR_{S1}}}\overset{2^{nR_{S1}}}{\underset{w_1=1}\sum}Pr[D_1^n(\bm
Y_1')\neq w_1|w_1 \,\, \mbox{is sent}]\rightarrow 0,\notag\\
&\underset{n\rightarrow \infty}\lim \left(  R_{S1} -
\frac{1}{n}H(W_1|\bm Y_2')\right)\rightarrow 0^{+}.\label{Def_Rs}
\end{align}

In the considered cognitive radio system, the primary user should
not be required to change his encoder and decoder due to the
presence of secondary users. Therefore, T$_1$ and U$_1$ are
restricted to use the usual wiretap encoder and decoder, which are
designed for a three-node wiretap channel: one legitimate
transmitter, one legitimate receiver, and one eavesdropper. The
secrecy rate $R_{S1}$ that can be achieved under this condition is
defined as a {\it single-user achievable secrecy rate}. On the other hand, T$_2$
intends to send an independent message $W_2\in
\mathcal{W}_2=\{1,\cdots,2^{n R_2}\}$ to the secondary receiver
U$_2$ in $n$ channel uses with the following encoding and decoding rules
\begin{align}
E_2^n:\, E_1^n\times \mathcal{W}_1\times \mathcal{W}_2\rightarrow
\mathcal{X}_2^n \mbox{ and }D_2^n:\, \mathcal{Y}_2'^n \rightarrow
\mathcal{W}_2.\notag
\end{align}


A rate $R_2$ is achievable for the secondary user if there exists a sequence of $(2^{nR_2},n)$ codes such that
\begin{align}
\underset{n\rightarrow \infty}\lim
\frac{1}{2^{nR_2}}\overset{2^{nR_2}}{\underset{w_2=1}\sum}Pr[D_2^n(\bm
Y_2')\neq w_2|w_2 \,\, \mbox{is sent}]\rightarrow 0,
\end{align}
while satisfying the secure
coexistence conditions.

\section{Transmission Schemes and Achievable Rate Regions}\label{sec:schemes}
In this section we
first discuss our main results which include the secondary user's
achievable rate in discrete memoryless channels (DMC) with
corresponding constraints to guarantee the primary user's security
and the extension to multi-phase transmission. Then we investigate different transmission schemes and
their corresponding achievable rate pairs in AWGN channels. We proposed a 3-phase clean relaying scheme combined with
dirty paper coding (DPC)\footnote{When considering AWGN channel, we
use dirty paper coding (DPC) to replace Gelfand-Pinsker Coding
(GPC), which is optimal for perfect channel state information at the
transmitter.} and cooperative jamming with numerical assessments in
the next section.

\subsection{3-Phase Clean Relaying Scheme with Gelfand-Pinsker Coding}
\label{sub:DPC} In this section we discuss the case where T$_2$ uses
GPC to precode T$_1$'s signal.

\subsubsection{Single Phase Transmission for DMC}
To guarantee the secrecy of the primary user's transmission when
T$_2$ exploits GPC, we need to derive the reliability and analyze
the equivocation rate of the achievable scheme, but not directly
apply GPC to precode the primary user's signal as the traditional
interference mitigation based cognitive radio without secrecy
constraint \cite{Jovicic_CR}. Denote the auxiliary random variables
transmitted by T$_1$ and T$_2$ as $V_1$ and $V_2$, respectively. The
main result is then given as follows. \vspace{-0.3cm}
\begin{theorem}\label{Th_conditions_single_binning_equals_double}
Assume T$_2$ non-causally knows $V_1^n$ and the primary user's
coding scheme is fixed with rate $R_{S1}\triangleq
I(V_1;Y_1)-I(V_1;Y_2)$ {and the rate per bin is $R_{S1}'\triangleq
I(V_1;Y_2)-2\epsilon$}. Then the rate $R_2$ is achievable such that
the secure coexistence conditions are satisfied, where
\begin{equation}
R_2=\max_{p_{V_2|V_1},\,p_{X_2|V_2,V_1}} I(V_2;Y_2')-I(V_2;V_1),
\end{equation}
\begin{numcases}{\hspace{2cm}\txm{   s.t.}}
   R_{S1}\leq I(V_1;Y_1')-R_{S1}', \label{EQ_old_new_main_channel}\\
   R_{S1}' = I(V_1;V_2,{Y_2'}^{}).\label{EQ_analysis_lambda2_constraint00}
  \end{numcases}
\end{theorem}

The proof is derived in
\ref{App_conditions_single_binning_equals_double}. The additional
constraints \eqref{EQ_old_new_main_channel} and
\eqref{EQ_analysis_lambda2_constraint00} can be explained
intuitively as follows: \eqref{EQ_old_new_main_channel} ensures that
once T$_2$ transmits, the equivalent main  channel can support the
transmission of the primary's code such that U$_1$ can successfully
decode all of the $2^{n[R_{S1}+R_{S1}']}$ codewords with negligible error,
while \eqref{EQ_analysis_lambda2_constraint00} guarantees the
secrecy, i.e., the wiretap code sent by T$_1$ to U$_2$ can
be supported by the new channel when T$_2$ transmits in the sense
that, given $W_1=w_1$, U$_2$ is able to find one codeword which is
jointly typical to the received signal. Therefore, U$_2$ will be
completely confused and the information leakage due to the use of
GPC at T$_2$ can be mitigated.

\begin{rmk}\label{RMK_comparison}
{Note that due to the secure coexistence conditions, the rate
expression in Theorem
\ref{Th_conditions_single_binning_equals_double} is different to
those in the cognitive interference channels
\cite{Liang2009}, \cite{Somayeh}. First, here condition
\eqref{EQ_analysis_lambda2_constraint00} is additionally required to
guarantee that the primary user's original secrecy rate
$I(V_1;Y_1)-I(V_1;Y_2)$ is achievable. Second, our rate expression
is composed of two channel models, i.e., the cases where either
T$_2$ transmits or not, which cannot be specialized from
\cite{Liang2009}, \cite{Somayeh}.}
\end{rmk}


\begin{rmk}
Note that we cannot simplify the constraint by substituting
\eqref{EQ_analysis_lambda2_constraint00} into
\eqref{EQ_old_new_main_channel}. The reason is as follows. Recall
that we describe the binning codebook of wiretap codes by two
dimensions (or indices): bin indices (or rows) denote the secured
message and the codewords in each bin (or columns) are to confuse
T$_2$. More specifically, the left hand side (LHS) of
\eqref{EQ_analysis_lambda2_constraint00} is the number of codewords
representing the same message, which is fixed at T$_1$. If we
simplify the constraint by doing so, we will have a constraint only
on the \textit{target} secrecy rate. However, the transmission at
that rate may not be secure in the weak sense. This is because
without such constraint, there is no enough codewords representing
the same message to confuse the eavesdropper T$_2$.
\end{rmk}

\begin{rmk}
Since single binning is used in the primary user's codebook and must
be left intact by the cognitive users, during the analysis of the
equivocation rate the derivation of $H(\bm V_1|\bm V_2, \bm
Y_2',W_1)$ need further analysis than cases in which $\bm V_1$ belong to
a double binning codebook when GPC is used, e.g.,
\cite{Chen_wiretap_SI}, \cite{Liu_SISO}. More specifically, in the
latter case, the remaining uncertainty of $\bm V_1$ given $\bm V_2$
can be simply derived by removing one of the sub bin indices. On the
contrary, here we need to derive a tight upper bound of $H(\bm
V_1|\bm V_2, \bm Y_2',W_1)$ to guarantee a larger cognitive user's rate. Please refer to Appendix
II for a more detailed
discussion.
\end{rmk}
\subsubsection{Three-Phase Transmission for DMC}
The motivations of the three-phase transmission scheme are twofold:
1) To provide an additional degree of freedom by different
transmission schemes to improve the secondary user's rate depending
on the location of nodes and their channels; 2) To take into account
the latency of successful decoding of $w_1$ at T$_2$ to accommodate
the practical scenarios where $w_1$ cannot be known non-causally at
T$_2$.

In the following we extend the result in Theorem
\ref{Th_conditions_single_binning_equals_double} to a three-phase
transmission, where the distribution of each phase can be
respectively factorized from \eqref{EQ_DMC1} as follows
\begin{align}
P_{{Y_1'}^{(1)}, \,{Y_2'}^{(1)},\,V_1} &=  \,P_{{Y_1'}^{(1)},\, {Y_2'}^{(1)}|V_1}P_{V_1},\notag\\
P_{{Y_1'}^{(2)},\, {Y_2'}^{(2)},\,X_2^{(2)},\, V_1, V_2} &= \, P_{{Y_1'}^{(2)},\,{Y_2'}^{(2)}|X_2^{(2)},\,V_1}P_{X_2^{(2)}|V_1,\,V_2}P_{V_2|V_1}P_{V_1},\notag\\
P_{{Y_1'}^{(3)},\, {Y_2'}^{(3)},\,X_2^{(3)},\, V_1} &=  \,
P_{{Y_1'}^{(3)},\,{Y_2'}^{(3)}|X_2^{(3)},\,V_1}P_{X_2^{(3)}|V_1}P_{V_1},\notag
\end{align}
where the notation $A^{(k)}$ denotes random variable $A$ in the
$k-$th phase. Note that distributions of $V_1$ and $V_2$ are fixed
over the three phases, thus we omit the specification of the phase
to simplify the notation.

Assume the time index $t\in \mathds{N}$. We define the sets of the
three phases as $\textsf{T}_1=\{t:1\leq t\leq\lfloor\eta_1
n\rfloor\}$, $\textsf{T}_2=\{t:\lfloor\eta_1 n\rfloor+1\leq
t\leq\lfloor(\eta_1+\eta_2) n\rfloor\}$,
$\textsf{T}_3=\{t:\lfloor(\eta_1+\eta_2) n\rfloor+1\leq t\leq n\}$,
respectively, with $\eta_1+\eta_2+\eta_3=1,\,0<\eta_1<1,\,0\leq\eta_2,
0\leq\eta_3$. Note that $\lfloor\eta_1 n\rfloor$ must be no less
than the time that T$_2$ needs to successfully decode $w_1$
\cite{Azarian_advance_decoding} and $\eta_1$ must be less than 1 for
T$_2$'s own transmission, i.e., a nonzero duration of the second
phase. The former condition can be modeled by letting T$_1$, T$_2$, and
U$_1$ form a degraded wiretap channel\footnote{Here we consider the
stochastic degradedness \cite{kim11}, i.e., there exists a random
variable $\tilde{Y}_T$ denoting the received signal at T$_2$ such that 1) $\tilde{Y}_T|\{X=x\}\sim
p_{\tilde{Y}_T|X}(\tilde{y}_T|x)$, and 2) $X\rightarrow
\tilde{Y}_T\rightarrow Y_1$ form a Markov chain.}. Besides, these
ratios are fixed before each transmission according to the
optimization results which will be discussed at the end of this
section. We want to design T$_2$'s transmit signal $\bm x_2$ where
the secure coexistence conditions are satisfied and we
consider the weak secrecy constraint.


\begin{proposition}\label{Th_conditions_single_binning_equals_double}
Assume that the primary user's coding scheme is fixed with code rate
$R_{S1}\triangleq I(V_1;Y_1)-I(V_1;Y_2)${, the rate per bin is
$R_{S1}'\triangleq I(V_1;Y_2)-2\epsilon$,} and T$_1$, T$_2$, and U$_1$
form a degraded wiretap channel. Then rate $R_2$ is achievable such
that the reliable decoding and the weak secrecy can be guaranteed where
\begin{equation}
R_2=\max_{p_{V_2|V_1},\,p_{X_2|V_2,V_1}} \eta_2\left(I(V_2;Y_2'^{(2)})-I(V_2;V_1)\right),
\end{equation}
\begin{numcases}{\hspace{2cm} \txm{   s.t.}}\,R_{S1}+R_{S1}'\leq \sum_{k=1}^3\eta_k I(V_1 ;\!Y_1'^{(k)}),\label{EQ_old_new_main_channel2}\\
R_{S1}'\!=\!\eta_1 I\!(V_1 ;\!Y_2'^{(1)}\!)\!+\!\eta_2
I\!(V_1;\!V_2,{Y_2'}^{(2)}\!)\!+\!\eta_3 I\!(V_1
;\!Y_2'^{(3)}\!).\label{EQ_analysis_lambda2_constraint002}
\end{numcases}
\end{proposition}
The result in Proposition
\ref{Th_conditions_single_binning_equals_double} can be derived by
applying the concept of parallel channels \cite{Liang_fading} such
that the achievable secrecy rate takes the average over the
considered three transmission phases in addition with the fact that
$V_2$ only exists in the second phase. Thus the proof is omitted.

\subsubsection{Three-phase Transmission for AWGN
Channels}\label{Sec_signaling}
The three-phase transmission for AWGN channels is explained as follows:\\
\textbf{Phase 1} ($t\in\textsf{T}_1$): The following decodability
constraint is required \cite{Azarian_advance_decoding}
\begin{align} \label{EQ_constraint_decodability}
|c_{TT}|>|c_{11}|,
\end{align}
which can be easily seen from the degradedness assumption of $Y_1$ and $Y_T$. In the following we normalize $|c_{11}|$ to 1. 

\noindent \textbf{Phase 2} ($t\in\textsf{T}_2$): the transmit power of T$_2$ in this phase $P_2^{(2)}$ is
divided into three parts:

\begin{enumerate}
\item \textbf{Jamming:} the jamming signal $a_{2}(t)$ has power $P_{2a}^{(2)} = \rho_2
P_2^{(2)}$ to confuse the eavesdropping secondary user U$_2$. The
parameter $\rho_2\in[0,1)$ denotes the fraction of the power used
for jamming.
\item \textbf{Relaying of the primary message:}
If \eqref{EQ_constraint_decodability} is valid, T$_2$ helps to relay
$\{x_1(t)\}_{t\in \textsf{T}_2}$ in Phase $2$ while simultaneously
transmitting its own message $w_2$. The relay power is
$P_{2,1}^{(2)} = \gamma(1-\rho_2) P_2^{(2)}$, where $\gamma$ is the
ratio of the remaining power for relaying.
\item \textbf{Transmission of the secondary message:} $w_2$ is encoded into
$\bm v_{2}^{(2)}$ with power $(1-\gamma)(1-\rho_2) P_2^{(2)}$ to be
decoded by U$_2$ only.
\end{enumerate}

\noindent \textbf{Phase 3} ($t\in \textsf{T}_3$): T$_2$ relays $\{x_1(t)\}_{t\in \textsf{T}_3}$ with power
{$(1-\rho_3)P_{2}^{(3)}$} and transmits the jamming signal with power $\rho_3
P_{2}^{(3)}$, but without super-imposing its own signal $v_2(t)$.


The average transmit power constraints for T$_k$, are
\begin{equation}
\frac{1}{n}\sum_{i=1}^{n}\left|x_{k}(i)\right|^{2} \leq P_k
\hspace{0.5cm} \text{for} \hspace{0.5cm} k \in \{1,2\}.
\end{equation}
More specifically, the transmit power constraint at T$_2$ is
\begin{align}
\eta_2 P_2^{(2)}+\eta_3 P_2^{(3)}\leq P_2,
\end{align}
where $
P_2^{(k)}\triangleq\frac{1}{|\textsf{T}_k|}\sum_{t\in\textsf{T}_k}\left|x_{2}(t)\right|^{2}
$, $k=2\mbox{ and }3$.

Assume that the noises at all nodes are independent and identically
distributed circularly symmetric complex AWGN with zero mean, unit
variance and are mutually independent for all $t$. We also assume
that T$_1$ perfectly knows the channel states from T$_1$ to U$_1$
and from T$_1$ to U$_{2}$, while T$_2$ knows all channel states.
{Note that when $|c_{TT}|\leq 1$, T$_2$ cannot know what is
transmitted due to the wiretap code used at T$_1$. On the contrary,
if $|c_{TT}|> 1$, T$_2$ can accomplish the successful decoding in
the same way as at the legitimate receiver U$_1$, since $|c_{TT}|$
is large enough for T$_2$ to decode all the codewords in the binning
codebook successfully.\footnote{The celebrated Shannon's random
codebook is adopted in our paper, which guarantees the existence of
codebook with decodable $w_1$ in Phase 1 as in the seminal papers
\cite{Mitran_ST}\cite{Azarian_DMT}. } }

When T$_{2}$ does not transmit, the secrecy capacity of the primary
user is $\left(\Ca(P_{1})-\Ca(|c_{12}|^{2}P_{1})\right)^{+}$. Define
\begin{align}
c_{11}^{(2)}&\triangleq
1+c_{21}e^{-j\phi_{21}}\sqrt{(1-\rho_2)\gamma
P_2^{(2)}/P_1},\,\,c_{12}^{(2)}\triangleq c_{12}+c_{22}e^{-j\phi_{21}}\sqrt{(1-\rho_2)\gamma
P_2^{(2)}/P_1},\\
c_{11}^{(3)}&\triangleq
1+c_{21}e^{-j\phi_{21}}\sqrt{(1-\rho_3) P_2^{(3)}/P_1},\,\,c_{12}^{(3)}\triangleq c_{12}+c_{22}e^{-j\phi_{21}}\sqrt{(1-\rho_3)
P_2^{(3)}/P_1},
\end{align}
where $\phi_{21}$ is the phase of $c_{21}$. The secondary user's
rate with DPC is given in the following proposition.
\begin{proposition}\label{propDPC}
Assume that the primary user's coding scheme is fixed with rate
$R_{S1}=\left(\Ca(P_1)-\Ca(|c_{12}|^2 P_1)\right)^+$. When DPC is used
at T$_2$, the rate $R_2$ is achievable such that the secure coexistence conditions can be guaranteed where
\begin{align}
R_2=\eta_2
\Ca\left(\frac{|c_{22}|^2P_{U_2}}{1+|c_{22}|^2\rho_2P_2^{(2)}}\right),
\end{align}
if \eqref{EQ_constraint_decodability} and the following constraints
\begin{align}
&\eta_2 \Ca\left(\frac{|c_{11}^{(2)}|^2P_1}{1+|c_{21}|^2(\rho_2 P_2^{(2)}+P_{U_2})}\right) +\eta_3 \Ca\left(\frac{|c_{11}^{(3)}|^2P_1}{1+|c_{21}|^2\rho_3 P_2^{(3)}}\right)\geq (1-\eta_1)\Ca(P_1), \label{EQ_AWGN_constraint1}\\
&\eta_2 \Ca\left(\frac{|c_{12}^{(2)}|^2P_1}{1+|c_{22}|^2(\rho_2
P_2^{(2)}+P_{U_2})}\right) +\eta_3
\Ca\left(\frac{|c_{12}^{(3)}|^2P_1}{1+|c_{22}|^2\rho_3
P_2^{(3)}}\right)\label{EQ_AWGN_constraint2}=
(1-\eta_1)\Ca(|c_{12}|^2P_1),
\end{align}
are fulfilled, where $P_{U_2}\triangleq(1-\rho_2)(1-\gamma)P_2^{(2)}$.
\end{proposition}
The proof is derived in \ref{App_DPC_AWGN}.

Based on Proposition
\ref{Th_conditions_single_binning_equals_double} and the fact that a
smaller $\eta_1$ results in a longer duration for transmitting the
secondary user's signal and clean relaying, we set
$\eta_1=\Ca(P_1)/\Ca(|c_{TT}|^2 P_1)$, where $\lfloor\eta_1
n\rfloor$ is the smallest duration for T$_2$ to successfully decode
T$_1$'s message. We then define the optimization problem for this
case as
\begin{equation}
\textbf{P1:}\,\,\max_{\eta_2,\,\rho_2,\,\rho_3,\,\gamma,\,P_2^{(2)}, P_2^{(3)}}  \ R_2\label{EQ_Optimization}
\end{equation}
\begin{numcases}
{\txm{s.t. }  \eqref{EQ_constraint_decodability},\,\eqref{EQ_AWGN_constraint1},\,\eqref{EQ_AWGN_constraint2} \txm{ and }}
 \eta_2 P_2^{(2)} + (1-\eta_1-\eta_2) P_2^{(3)} \leq P_2,\,P_2^{(2)}\geq 0,\,P_2^{(3)}\geq 0 \label{power_constraint}\\
 \rho_2\geq 0,\,\rho_3\geq 0,\,\gamma\geq 0 \label{constraint_rho}\\
 \eta_2+\eta_3=1-\eta_1,\,\eta_2\geq 0,\,\eta_3\geq 0,
\label{EQ_power_constraint}
\end{numcases}
where the first inequality in \eqref{power_constraint} is the average power constraint. Instead of
focusing on solving the non-convex $\textbf{P1}$, we will thoroughly
investigate it with numerical illustrations in Section
\ref{sec:numericalResults} and investigate the impact of the
optimization variables on the rates for relevant scenarios.

\begin{rmk}\label{RMK_no_CJ}
{Note that due to the secure coexistence condition (i), pure
cooperative jamming cannot be used here. This is because, there are
$2^{n (I(V_1;Y_1)-\epsilon)}$ codewords in the  primary user's
codebook which is fixed, and U$_1$ must be able to successfully
decode all the codewords. When pure CJ is used, the capacity of the
main channel becomes $I(V_1;Y_1')$ which is smaller than
$I(V_1;Y_1)$. Then the reliability constraint of the transmission to
the legitimate receiver T$_1$ is invalid.}
\end{rmk}

\begin{rmk}\label{RMK_conjecture}
It is interesting to see whether transmitting relay and jamming signals separately by different phases improves the performance or not, comparing to the proposed 3-phase scheme where both signals are transmitted in the third phase. We conjecture that the proposed 3-phase scheme does not perform worse than schemes with 4-phase. We first sketch the steps to identify this question as follows. Firstly, we extend the optimization problem $\mathbf{P1}$ into four phases to encompass phases for pure relay and pure jamming. By comparing $R_1$ solving from three and four-phase schemes numerically, we can observe that only one of the third and forth phases of the 4-phase scheme is selected. This means the proposed 3-phase scheme performs the same as the the four-phase one under the considered scenario. The detailed derivation is as follows.
Due to the intractability of the optimization
problems, we mainly show the performance of the 4-phase
scheme numerically complemented with thorough qualitative discussions. The DMC rate of the cognitive user with a 4-phase transmission scheme can be expressed as

\begin{equation}
\hspace{-5cm}R_2=\max_{p_{V_2|V_1},\,p_{X_2|V_2,V_1}} \eta_2\left(I(V_2;Y_2'^{(2)})-I(V_2;V_1)\right),
\end{equation}
\begin{numcases}{\hspace{2cm} \txm{   s.t.}}\,R_1+R_1'\leq \sum_{k=1}^4\eta_k I(V_1 ;\!Y_1'^{(k)}),\label{EQ_4PH_constr1}\\
R_1'\!=\!\eta_1 I\!(V_1 ;\!Y_2'^{(1)}\!)\!+\!\eta_2
I\!(V_1;\!V_2,{Y_2'}^{(2)}\!)\!+\!\eta_3 I\!(V_1
;\!Y_2'^{(3)}\!)+\!\eta_4 I\!(V_1
;\!Y_2'^{(4)}\!).\label{EQ_4PH_constr2}
\end{numcases}

To compare schemes with 3- and 4-phases in AWGN cases, we first evaluate the terms in the third and fourth phases in \eqref{EQ_4PH_constr1} and \eqref{EQ_4PH_constr2} as follows, where we assume that in the third phase we only relay but do not transmit jamming signals and in the fourth phase we only transmit the jamming signals but do not relay:

\begin{align}
\eta_3' I(V_1 ;\!Y_1'^{(3)})+\eta_4 I(V_1 ;\!Y_1'^{(4)})&=
\eta_3'\Ca(|c_{11}^{(3)}(0,P_2'^{(3)})|^2P_1)+\eta_4\Ca\left(\frac{P_1}{1+|c_{21}|^2P_2^{(4)}}\right),\label{EQ_4PH_constr1_AWGN}\\
\eta_3' I(V_1 ;\!Y_2'^{(3)})+\eta_4 I(V_1 ;\!Y_2'^{(4)})&=
\eta_3'\Ca(|c_{12}^{(3)}(0,P_2'^{(3)})|^2P_1)+\eta_4\Ca\left(\frac{|c_{12}|^2P_1}{1+|c_{22}|^2P_2^{(4)}}\right),\label{EQ_4PH_constr2_AWGN}
\end{align}

where $c_{11}^{(3)}(\rho,P)\triangleq 1+c_{21}e^{-j\phi_{21}}\sqrt{\frac{(1-\rho)P}{P_1}}$ and $c_{12}^{(3)}(\rho,P)\triangleq c_{12}+c_{22}e^{-j\phi_{21}}\sqrt{\frac{(1-\rho)P}{P_1}}$ are equivalent channels; $P_2'^{(3)}$ is the CR power used in the third phase of the 4-phase scheme; $P_2^{(3)}$ is the CR power used in the third phase of the 3-phase scheme.

Similarly, we use $\eta_3'$ to distinguish the third phase of the 4-phase scheme from $\eta_3$ of the 3-phase scheme. Recall that $\rho_3$ is the power ratio for jamming in the third phase. We also restate the third phase in the corresponding constraints in the 3-phase scheme as follows for the convenience of comparison
\begin{align}
\eta_3 I(V_1 ;\!Y_1'^{(3)})&=
\eta_3\Ca\left(\frac{|c_{11}^{(3)}(\rho_3,P_2^{(3)})|^2P_1}{1+|c_{21}|^2\rho_3P_2^{(3)}}\right),\label{EQ_3PH_constr1_AWGN}\\
\eta_3 I(V_1 ;\!Y_2'^{(3)})&=
\eta_3\Ca\left(\frac{|c_{12}^{(3)}(\rho_3,P_2^{(3)})|^2P_1}{1+|c_{22}|^2\rho_3P_2^{(3)}}\right).\label{EQ_3PH_constr2_AWGN}
\end{align}
From the above we can express the primary user's sum secrecy rate of the third phase and the fourth phase by the 4-phase scheme as
\begin{align}\label{EQ_R1_4PH}
R_{1,4PH}^{(3,4)}=&\eta_3'\left\{\Ca(|c_{11}^{(3)}(0,P_2'^{(3)})|^2P_1)-\Ca(|c_{12}^{(3)}(0,P_2'^{(3)})|^2P_1)\right\}+\notag\\
&\eta_4\left\{\Ca\left(\frac{P_1}{1+|c_{21}|^2P_2^{(4)}}\right)-\Ca\left(\frac{|c_{12}|^2P_1}{1+|c_{22}|^2P_2^{(4)}}\right)\right\}.
\end{align}
And the primary user's rate of the last phase, i.e., the third phase, by the 3-phase scheme is restated as
\begin{align}\label{EQ_R1_3PH}
R_{1,3PH}^{(3)}=\eta_3\left\{\Ca\left(\frac{|c_{11}^{(3)}(\rho_3,P_2^{(3)})|^2P_1}{1+|c_{21}|^2\rho_3P_2^{(3)}}\right)-
\Ca\left(\frac{|c_{12}^{(3)}(\rho_3,P_2^{(3)})|^2P_1}{1+|c_{22}|^2\rho_3P_2^{(3)}}\right)\right\}.
\end{align}

To compare the optimal $R_{1,4PH}^{(3,4)}$ and $R_{1,3PH}^{(3)}$ under the same parameters in the first and second phases, we need to impose the following two constraints
\begin{align}
\eta_3'P_2'^{(3)}+\eta_4P_2^{(4)}&=\eta_3P_2^{(3)},\label{EQ_new_constr1}\\
\eta_3'+\eta_4=\eta_3&=1-\eta_1-\eta_2,\label{EQ_new_constr2}
\end{align}
where $0<\eta_1<1$ and $0<\eta_2<1$ are predefined and fixed.

Then we check that whether the additional fourth phase in \eqref{EQ_R1_4PH} provides a performance gain, i.e.,
\begin{align}\label{EQ_whether_4PH_is_better_than_3PH}
(R_{1,4PH}^{(3,4)})^*\overset{?}\gtrless (R_{1,3PH}^{(3)})^*,
\end{align}
where $(R_{1,3PH}^{(3)})^*=\max_{\rho_3,P_2^{(3)}} R_{1,3PH}^{(3)},\,s.t.\,\eqref{EQ_new_constr1},\eqref{EQ_new_constr2}$ and $(R_{1,4PH}^{(3,4)})^*=\max_{\eta_3',P_2^{(3)}} R_{1,4PH}^{(3,4)},\,\,s.t.\,\,\eqref{EQ_new_constr1},\,\eqref{EQ_new_constr2}$.

It is clear that \eqref{EQ_R1_4PH} and \eqref{EQ_R1_3PH} are not convex and it is hard to have an analytical solution. We resort to numerical method to answer \eqref{EQ_whether_4PH_is_better_than_3PH}. After enumerating different locations of the nodes and different transmit power constraints, we find that $(R_{1,4PH}^{(3,4)})^*= (R_{1,3PH}^{(3)})^*$. From the numerical result we observe that the optimized rates of the two schemes are the same. In addition, we can find that $\eta_3'$ can only be 0 or $\eta_3$. In particular, in the part above the blue line, $\eta_3'=\eta_3$ and below the blue line, $\eta_3'=0$. That means in total there will be always only three phases. Therefore, the proposed 3-phase scheme shows the same performance as always the 4-phase scheme. From numerical results we also observe that the achievable rates of the two schemes are identical.

Even though the above discussion is not a rigorous proof, based on it we reasonably conjecture that the 4-phase scheme does not outperform the proposed one.

\end{rmk}

\subsection{3-Phase Clean Relaying scheme without GPC}
\label{sub:CR} We can
easily specialize our previous result to the scenario where T$_2$ does not use GPC. In this section we
will show that a stronger secrecy measure than the commonly used one, i.e., the {variational distance}
\cite{Bloch_strong_secrecy} defined by $\sup|P_{W\bm Y_2}-P_{W}P_{\bm
 Y_2}|\leq\epsilon$ can be achieved without rate loss at the secondary user compared to the weak
 secrecy. We first derive the three-phase secrecy rate under this
 stronger secrecy measure.

\begin{theorem}\label{Thorem_3phase_gaussian_capacity}
For the three-phase transmission, the secrecy capacity of the discrete
memoryless wiretap channel under variational distance constraint can be
represented as
\begin{align}\label{EQ_3phase_Cs}
C_s=\sup_{\{( U^{(k)}, X_1^{(k)})\}\in \mathcal{P}} \sum_{k=1}^3
\eta_k \left\{I( U^{(k)}; Y_1^{(k)})-I( U^{(k)}; Y_2^{(k)})\right\},
\end{align}
where $\mathcal{P}\triangleq \{( U^{(k)},\, X_1^{(k)}):
U^{(k)}\rightarrow X_1^{(k)}\rightarrow  Y_1^{(k)} Y_2^{(k)}$ forms
a Markov chain, $k=1,\,2,\,3$ and $ \frac{1}{n}\Sigma_{j=1}^n
|x_j|^2\leq P_1\}$.
\end{theorem}
The proof of Theorem \ref{Thorem_3phase_gaussian_capacity} is given
in \ref{App_CT}. 


In the following we consider the 3-phase clean relaying scheme combined with
cooperative jamming as described in Section \ref{Sec_signaling}.
From Theorem \ref{Thorem_3phase_gaussian_capacity} we can find that
the enhancement of the secrecy level does not cause any loss of
secrecy capacity.

\begin{proposition}\label{propCR}
Assume the primary user's coding scheme is fixed with rate
$R_{S1}=\left(\Ca(P_1)-\Ca(|c_{12}|^2 P_1)\right)^+$. Without using DPC
at T$_2$, the rate $R_2$ is achievable while the reliable
decoding and the secrecy measure of variational distance can be guaranteed, where
\begin{align}
R_{2} \leq \max&   \,\,\eta_2\,\Ca\left(\frac{|c_{22}|^2(1-\rho_2)
(1-\gamma)
P_2^{(2)}}{1+|c_{22}|^2\rho_2P_2^{(2)}+\left|c_{12}^{(2)}\right|^2P_1}
\right)\label{EQ_R2_ind_code}\\
s.t.& \,\,\Vast(
\eta_2\left\{\Ca\left(\frac{\left|c_{11}^{(2)}\right|^{2}P_1}{1 +
|c_{21}|^2 (1  - \gamma +  \gamma \rho_2) P_2^{(2)}} \right)- \Ca
\left( {\frac{\left|c_{12}^{(2)}\right|^{2}P_1} {1  +  |c_{22}|^2\rho_2 P_2^{(2)}}}\right)\right\}\notag\\
&\hspace{+1.21cm}+\eta_3\left\{ \Ca  \left(
\frac{\left|c_{11}^{(3)}\right|^{2}P_1}{1+|c_{21}|^2\rho_3
P_2^{(3)}}\right)- \Ca \left(
{\frac{\left|c_{12}^{(3)}\right|^{2}P_1}{1 +  |c_{22}|^2\rho_3
P_2^{(3)}}}   \right)  \right\} \Vast) ^{+} \geq(1-\eta_1)
R_{S1}.\label{EQ_R1_constraint_no_DPC}
\end{align}
\end{proposition}

The proof of Proposition \ref{propCR} follows the steps from Appendix III and is therefore omitted here. Without DPC, the optimization problem $\textbf{P1}$ simplifies into $\textbf{P2}$ as:
\begin{align}
\textbf{P2:}\,\,\max_{\eta_2,\,\rho_2,\,\rho_3,\,\gamma,\,P_2^{(2)}, P_2^{(3)}} & \ R_2\,\,\,\,\,\,\,\,
\mbox{s.t.
}\eqref{EQ_constraint_decodability},\,\eqref{power_constraint},\,\eqref{constraint_rho},\,\eqref{EQ_power_constraint}.\notag
\end{align}

\section{Upper Bound Analysis}\label{sec:UB}
{In this section we derive an upper bound for the channel in Fig. \ref{fig:2}. In particular, we first analyze the discrete memoryless channels in Section \ref{sec_UB_DMC} and then extend it to the AWGN channels in Section \ref{Sec_UB_AWGN}.

\subsection{Upper Bound for Discrete Memoryless Channels}\label{sec_UB_DMC}
The upper bound of the considered channel for discrete memoryless cases is given in the following theorem:

\begin{theorem}\label{Th_DMC_UB}
If the signal received at U$_2$ is a degraded version to that at U$_1$, the capacity-equivocation region outer bound is given by
\begin{equation}
\mathcal{C}_o^{DMC}=\bigcup_{\begin{subarray}{ll}U\rightarrow X_1\rightarrow Y_1\rightarrow Y_2,\\(U,V)\rightarrow X_2\rightarrow (Y_1,Y_2)\end{subarray}}\\
\left\{\begin{array}{l}
(R_1,R_2,R_{e_1})\in\mathds{R}^3_+:\notag\\
 R_{e_1}\leq R_1\notag\\
 R_1\leq I(U;Y_1)\notag\\
 R_2\leq I(V;Y_2)\notag\\
 R_1+R_2\leq I(X_1;Y_1|V)+I(V;Y_2)\notag\\
 R_{e_1}\leq I(X_1;Y_1|V)-I(X_1;Y_2|V)\end{array}\right\}.\notag
\end{equation}
\end{theorem}
\vspace{0.5cm}
The proof is derived in \ref{App_DMC_UB}.

The secrecy rate upper bound of the primary user can be expressed as
\begin{align}
R_{S1}=\min\{R_{e1},\,R_1\},
\end{align}
where
\begin{align}
R_1\leq \min\{I(U;Y_1), I(X_1;Y_1|V)+I(V;Y_2)\},
\end{align}
which can be easily derived from Theorem 1. In addition, since $I(X_1;Y_2|V)>0$, we know that \begin{align}\label{EQ_RS1_DMC}
R_{S1}=\min\{I(X_1;Y_1|V)-I(X_1;Y_2|V),\,I(U;Y_1)\}.
\end{align}
Meanwhile, the upper bound of $R_2$ can be derived as\footnote{With abuse of notation, here we use $R_{S1,\,target}$ to denote the primary user's target secrecy rate, which has the same meaning as the one in \eqref{Def_Rs}, but is different to the one in \eqref{EQ_RS1_DMC}.}
\begin{align}\label{EQ_R2_selection}
R_2&\leq \min\{I(V;Y_2),\, I(X_1;Y_1|V)+I(V;Y_2)-R_{S1,\,target}\}=I(V;Y_2) - (R_{S1,\,target}-I(X_1;Y_1|V))^+.
\end{align}

\subsection{Upper Bound for AWGN Channels}\label{Sec_UB_AWGN}
In the following we derive the capacity outer bound for AWGN
channels.

\begin{theorem}\label{Th_AWGN_UB}
If the signal received at U$_2$ is a degraded version to that at U$_1$, the outer bound of the rate pair $(R_{s1},R_2)$ is given by
\begin{equation}
(R_{s1},R_2)=\bigcup_{\begin{subarray}{l}0\leq\alpha,\beta,\delta,\eta,\gamma\leq 1,\notag\\
|\rho|\leq 1\end{subarray}} \left\{\begin{array}{l}
R_{s1}\leq\min\Bigg\{ \log\left(\frac{1+\tilde{P}_1+|a|^2\tilde{P}_2+2\Re\{a\rho\}\sqrt{\tilde{P}_1\tilde{P}_2}}{1+\alpha(\tilde{P}_1+|a|^2\tilde{P}_2+2\Re\{a\rho\}\sqrt{\tilde{P}_1\tilde{P}_2})}\right),\notag\\
\hspace{1.9cm} \Bigg(\log\left(\frac{1+\gamma(\tilde{P}_1+|a|^2\tilde{P}_2+2\Re\{a\rho\}\sqrt{\tilde{P}_1\tilde{P}_2})}
{1+\beta(|b|^2\tilde{P}_1+\tilde{P}_2+2\Re\{b\rho\}\sqrt{\tilde{P}_1\tilde{P}_2})}\right)-\log\left(\frac{1+\eta |a|^2 \tilde{P}_2}{1+\delta  \tilde{P}_1}\right)\Bigg)^+\Bigg\}\\
R_2\leq  \Bigg\{\log\left(\frac{1+|b|^2\tilde{P}_1+\tilde{P}_2+2\Re\{b\rho\}\sqrt{\tilde{P}_1\tilde{P}_2}}{1+\beta(|b|^2\tilde{P}_1+\tilde{P}_2+2\Re\{b\rho\}\sqrt{\tilde{P}_1\tilde{P}_2})}\right)-\notag\\
\hspace{1.2cm}\Bigg(\log\left(\frac{1+\tilde{P}_1}{1+|b|^2\tilde{P}_2}\right)-\log\left(\frac{1+\gamma(\tilde{P}_1+|a|^2\tilde{P}_2+2\Re\{a\rho\}\sqrt{\tilde{P}_1\tilde{P}_2})}{1+\eta |a|^2 \tilde{P}_2}\right)\Bigg)^+\Bigg\}
\end{array}\right\},\notag
\end{equation}
where $\tilde{P}_1\triangleq |H_{11}|^2P_1$, $\tilde{P}_2\triangleq |H_{22}|^2P_2$, $\rho\triangleq E[H_{11}{X}_1(H_{22}{X}_2)^*]/\sqrt{\tilde{P}_1\tilde{P}_2}$ is the correlation
coefficient.
\end{theorem}

The proof is derived in \ref{App_AWGN_UB}.

To compare the above UB to our existing lower bound, we need the following steps. To find the upper bound of the cognitive user's rate, we set the upper bound of $R_{S_1}$, namely $R_{S_1}^O$, to be the same as the primary user's target secrecy rate, namely, $R_{S_1,\,target}$, due to the secure coexistence condition (i). From numerical results we can find a set of $(\alpha,\beta,\delta,\eta,\gamma,\rho)$, namely, $\mathcal{S}$ such that the equality $R_{S_1}^O=R_{S_1,\,target}$ is valid. Then we can find the outer bound of the cognitive radio user's rate as
\begin{align}
&R_2^{o}= \max_{(\alpha,\beta,\delta,\eta,\gamma,\rho)\in\mathcal{S}}\log\left(\frac{1+|b|^2\tilde{P}_1+\tilde{P}_2+2\Re\{b\rho\}\sqrt{\tilde{P}_1\tilde{P}_2}}{1+\beta(|b|^2\tilde{P}_1+\tilde{P}_2+2\Re\{b\rho\}\sqrt{\tilde{P}_1\tilde{P}_2})}\right)-\notag\\
&\hspace{3.4cm}\Bigg(\log\left(\frac{1+\tilde{P}_1}{1+|b|^2\tilde{P}_2}\right)-\log\left(\frac{1+\gamma(\tilde{P}_1+|a|^2\tilde{P}_2+2\Re\{a\rho\}\sqrt{\tilde{P}_1\tilde{P}_2})}{1+\eta |a|^2 \tilde{P}_2}\right)\Bigg)^+.\end{align}
}
\vspace{0.5cm} 

\section{Numerical Illustrations}\label{sec:numericalResults}
In this section we investigate thoroughly the optimization problems
$\textbf{P1}$ and $\textbf{P2}$ using numerical illustrations based
on a relevant scenario. We first describe our geometrical setup in
Section \ref{sec:num_geo}. We then analyze the performance of our
3-phase clean relaying scheme, with and without DPC encoding at T$_2$ in
Section \ref{sec:num_perf}. After that we study the influence of the
optimization parameters to highlight some interesting behavior in
Section \ref{sec:num_param}. Finally we compare the achievable rate and the capacity upper bound of the secondary user.

\subsection{Setup}
\label{sec:num_geo} For our numerical illustrations of the
theoretical results, we are interested in the system behavior for
different locations of the secondary transmitter. In particular we
fix the locations of the primary transmitter T$_1$ and receiver
U$_1$ at the coordinates $(0,0)$ and $(1,0)$, respectively. The
secondary receiver is fixed at $(1,-1)$. We assume
 a path-loss model with path-loss exponent $\alpha = 3$, i.e., $c_{ij} = d_{ij}^{-3}$.
 The power constraints at T$_1$ is $P_1 = 10$ dB while we will consider different
 transmission powers at T$_2$ in order to measure the impact of
 the power constraint. Note that we also include power control as a possible design
parameter for T$_2$, i.e., the transmission power utilized is not
necessarily fixed to its maximum $P_2$. The unit of rate results is
bit per channel use. Note that we use the solver \textit{fmincon} from Matlab to solve the optimization problem numerically. Because fmincon is a derivative-based search algorithm, it cannot guarantee a global optimal solution for the considered problems, which are non-convex.

\subsection{Performance Analysis}

\begin{figure}
     \centering
     \resizebox{0.99\textwidth}{!}{
     \subfloat[width=0.45\textwidth][]{\includegraphics{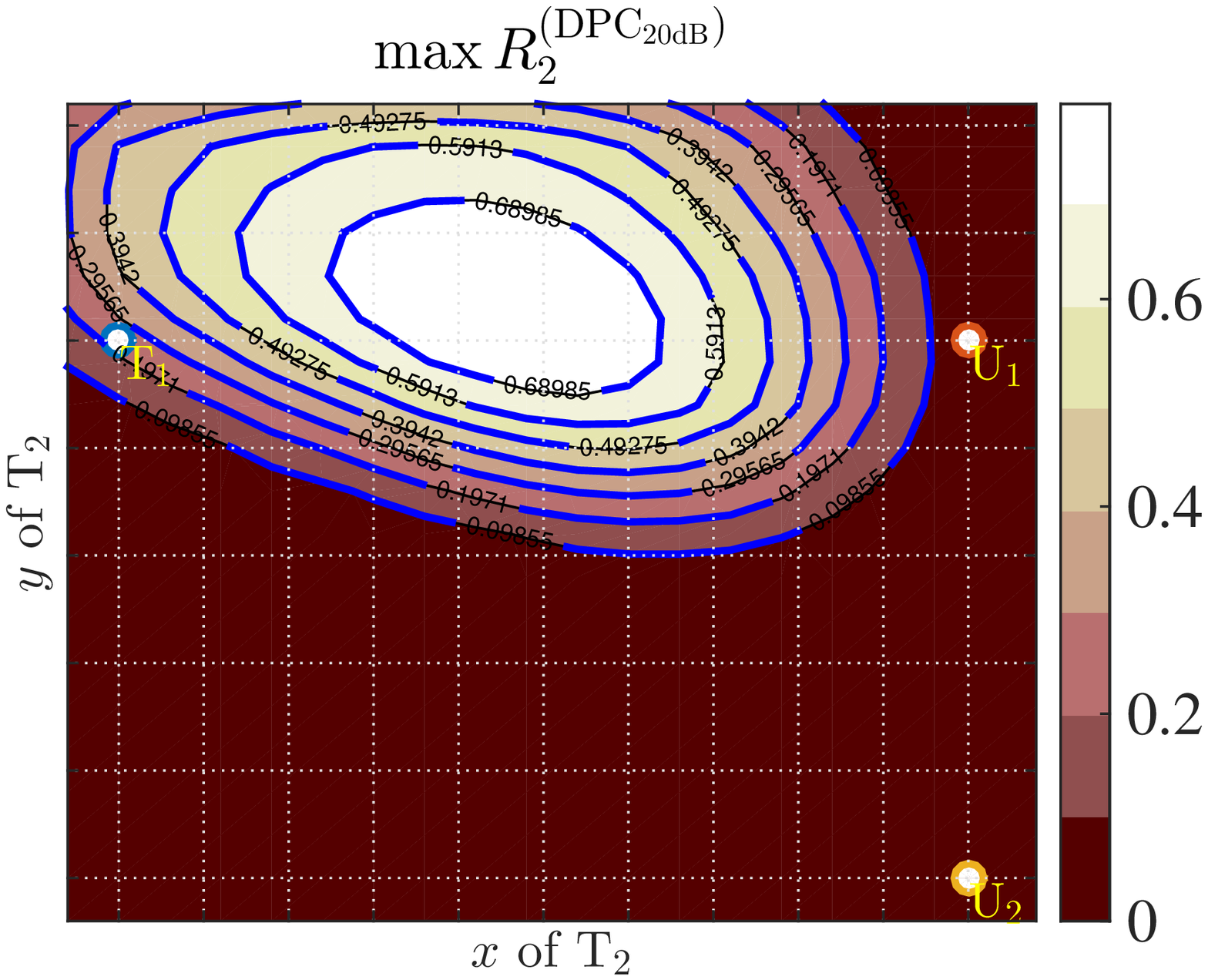}\label{fig:R2_20dB}}
     \subfloat[width=0.45\textwidth][]{\includegraphics{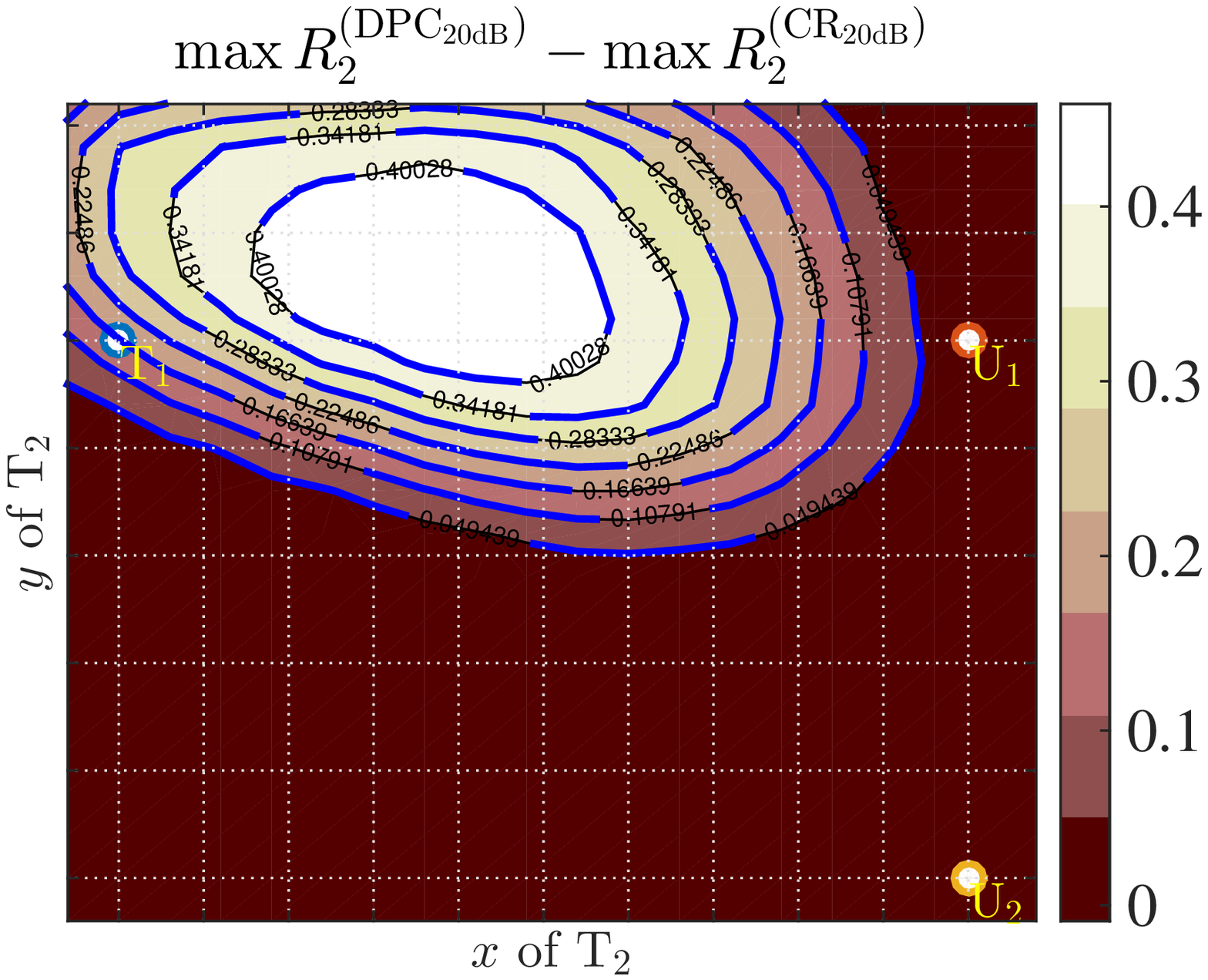}\label{fig:R2_diff_DPC_CR1}}}
     \caption{(a) Maximum achievable secondary user's rates
$R_2^{(\text{DPC}_{20\text{dB}})}$ as a function of the position of
T$_2$ with $P_2=20$ dB. (b)Difference between maximum achievable secondary user's
rates with and without DPC ($R_2^{(\text{DPC}_{20\text{dB}})}$ and
$R_2^{(\text{CR}_{20\text{dB}})}$) as a function of the position of
T$_2$ with $P_2=20$ dB.}
     \label{steady_state}
\end{figure}

\label{sec:num_perf}
\begin{enumerate}
\item{\textit{Performance of the 3-phase clean relaying scheme with DPC ($\textbf{P1}$)}:}

In this section we investigate the performance of the proposed
scheme. First we depict in Fig. \ref{fig:R2_20dB} the maximum
achievable secondary user's rates $R_2$ for the problem
$\textbf{P1}$, i.e., for the 3-phase clean relaying scheme combined with DPC
at T$_2$. In all figures, the $x$-axis combined with the $y$-axis
represents the coordinate of T$_2$ in the plane. The brighter the
point $(x,y)$ is, the higher the value of the depicted variable is
for this location of T$_2$. We observe from Fig. \ref{fig:R2_20dB}
that the highest secondary user's rates are attained for T$_2$
located between T$_1$ and U$_1$. Interestingly the highest rates are
not obtained for T$_2$ located close to U$_2$ which can be explained
in two ways. First due to the decodability condition, the location
of T$_2$ is bounded inside the circle around T$_1$ with radius as
the distance between T$_1$ and U$_1$. Second, since T$_2$ must
maintain the secrecy rate unchanged, it is primordial that T$_2$
helps the primary transmission in the classical relaying sense, for
which the optimal location of T$_2$ is between T$_1$ and U$_1$.

\item{\textit{Comparison with 3-phase clean relaying scheme without DPC ($\textbf{P2}$)}:}

In order to evaluate the gain from combining DPC with 3-phase clean
relaying scheme, we depict in Fig. \ref{fig:R2_diff_DPC_CR1} the difference
of the maximum achievable secondary user's rates achieved between
the optimizations $\textbf{P1}$ and $\textbf{P2}$. We observe that
the 3-phase clean relaying scheme with DPC provides gains up to 140$\%$ comparing
with scheme without DPC, especially when T$_2$ is located closer to
T$_1$.

\item{\textit{Comparison with the baseline scheme (without the third phase)} {and a 4-phase scheme}:}

\begin{figure}
\centering
\includegraphics[width=0.6\textwidth]{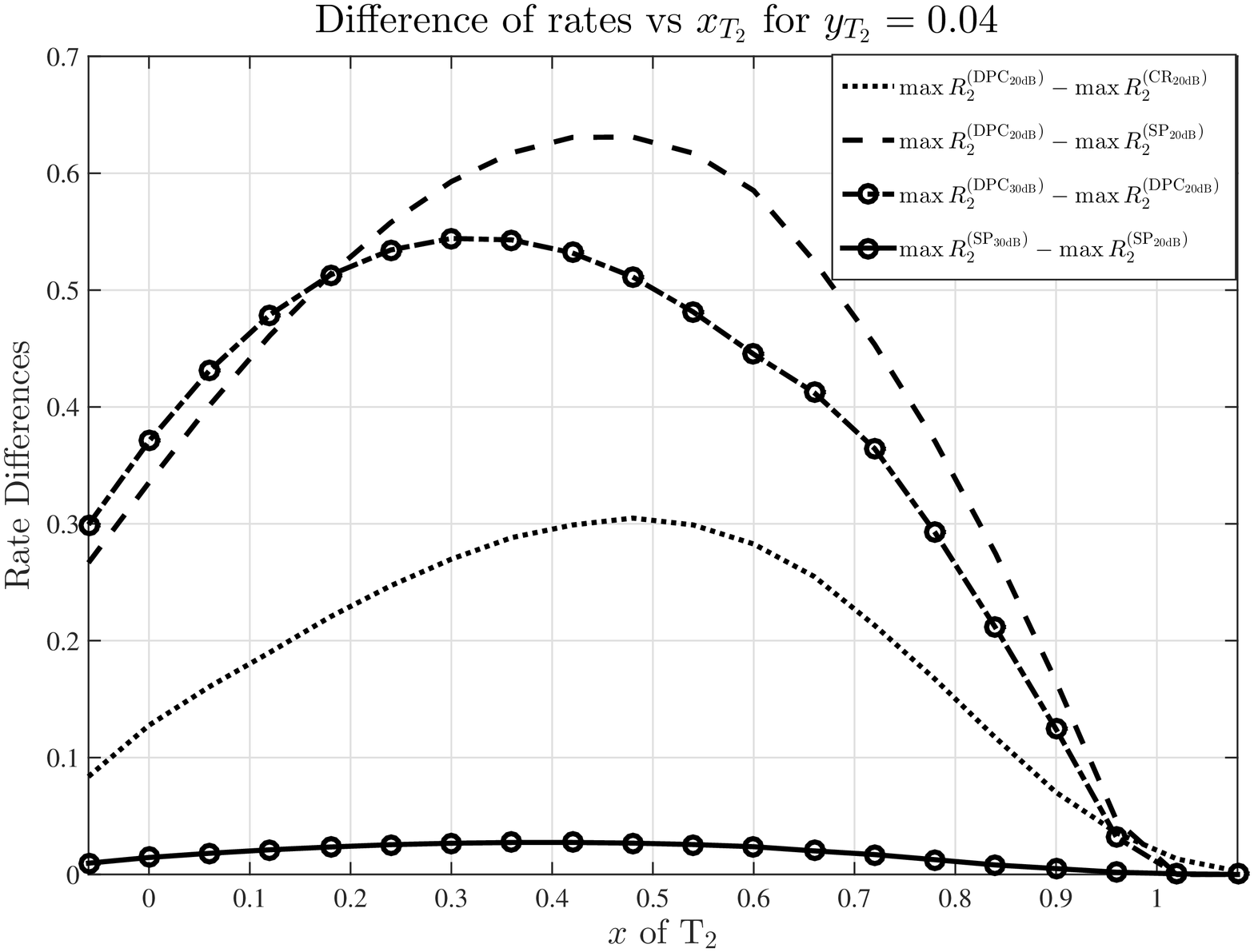}
\caption{Difference of maximum achievable secondary user's rates
between different schemes and different transmit powers as a function of the position of T$_2$.}\label{fig:R2_diff_scheme}
\end{figure}

%

\end{enumerate}

In Fig. \ref{fig:R2_diff_scheme} we compare the 3-phase clean
relaying scheme with DPC to the baseline scheme with the single-phase
(labeled as SP in the figure) transmission. Here single-phase is
defined as the one without clean relaying but still has the first
phase for listening and decoding the primary message. As expected,
the secondary user's rates obtained from the optimization
$\textbf{P1}$ are no worse than the rates without the clean relaying
phase, which was expected since the single-phase transmission scheme
is a special case of the 3-phase clean relaying scheme with $\eta_3=0$. More
specifically, from Fig. \ref{fig:R2_diff_scheme} we can observe that
the gain becomes significantly higher when T$_2$ is located between
T$_1$ and U$_1$.
 This observation is consistent to the intuition since when T$_2$ is at such location, the clean relaying in Phase 3 is highly efficient:
it will boost the signal to interference and noise ratio (SINR) of
$I(V_1;Y_1^{(3)})$ much faster than the relaying in Phase 2 because
T$_2$ is close to U$_1$. More specifically, under this geometry, in
Phase 2 the signal $U_2$ causes more interference to U$_1$, and thus
degrades the efficiency of relaying. Meanwhile the increment on the
SINR of $I(V_1;Y_2^{(3)})$ is limited due to T$_2$ being far enough
from U$_2$. Also in the same region, the power allocated to the
jamming signal can be reduced also due to T$_2$ being far enough
from U$_2$, which further improves the performance in this region.

{In this subsection we also compare the proposed three-phase scheme to a four-phase scheme in which the relayed and jamming signals are separated in the third and fourth phases, respectively. The characterization of the four-phase scheme can be straightforwardly extended from our three-phase scheme. Therefore, we omit the derivation here. By this comparison we aim to verify that the latter one does not provide any additional degree of freedom to improve $R_2$. From numerical result observe that $\eta_3$ can only be 0 or 1. In particular, in the part above the blue line, $\eta_3=1$ and below the blue line, $\eta_3=0$. That means in total there will be only three phases. Therefore, the proposed three-phase scheme is no worse than the four-phase scheme for the considered setup. }


\subsection{Relation between the Optimized Parameters and Geo-locations}\label{sec:num_param}
In this section we investigate our transmission schemes by illustrating some optimized parameters with respect to geo-locations. In particular we discuss the influence of two important parameters, namely the maximal transmission power of the secondary transmitter $P_2$ and the interval of the third phase $\eta_3$.
\begin{enumerate}
\item{\textit{Influence of the secondary power $P_2$}:}
We first investigate the influence of the maximal transmission power
of T$_2$. In particular we study the 3-phase clean relaying scheme with DPC and
single-phase transmission in Fig. \ref{fig:R2_diff_scheme} with
$P_2=20$dB and $P_2=30$dB. While we observe a somewhat similar
geographic behavior for both schemes, there is an important
difference between them. Indeed the magnitude of the difference
$R_2^{(\text{DPC}_{30\text{dB}})}-R_2^{(\text{DPC}_{20\text{dB}})}$
is significantly higher than that of the difference
$R_2^{(\text{SP}_{30\text{dB}})}-R_2^{(\text{SP}_{20\text{dB}})}$.
This behavior shows that higher secondary transmission power are
still beneficial when the 3-phase clean relaying scheme with DPC is used, while
increasing the transmission power without the clean relaying phase
is not efficient to enhance the rate.

\item{\textit{Influence of the third phase $\eta_3$}:}
\begin{figure}
\centering
\includegraphics[width=0.6\textwidth]{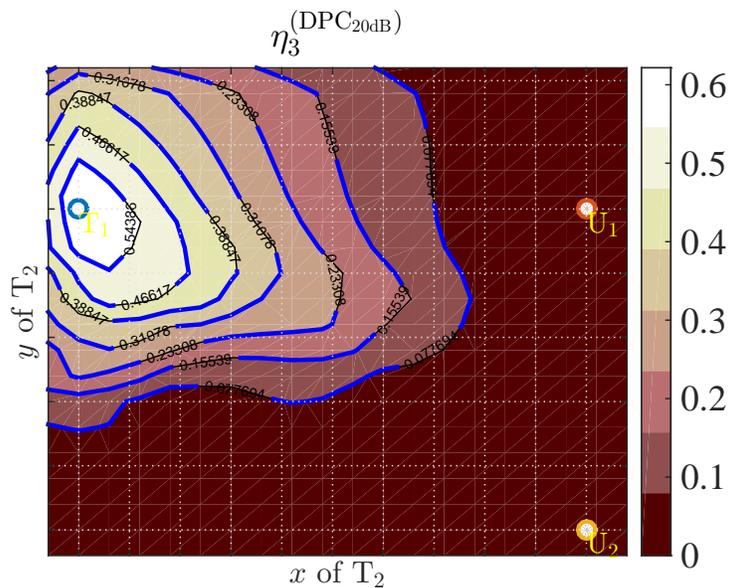}
\caption{Optimal
$\eta_3^{(\text{DPC}_{20\text{dB}})}$ for 3-phase clean relaying scheme with DPC
($\textbf{P1}$) with $P_2=20$dB  as a function of the position of
T$_2$.}\label{fig:eta3_DPC}
\end{figure}

\begin{figure}
\centering
\includegraphics[width=0.6\textwidth]{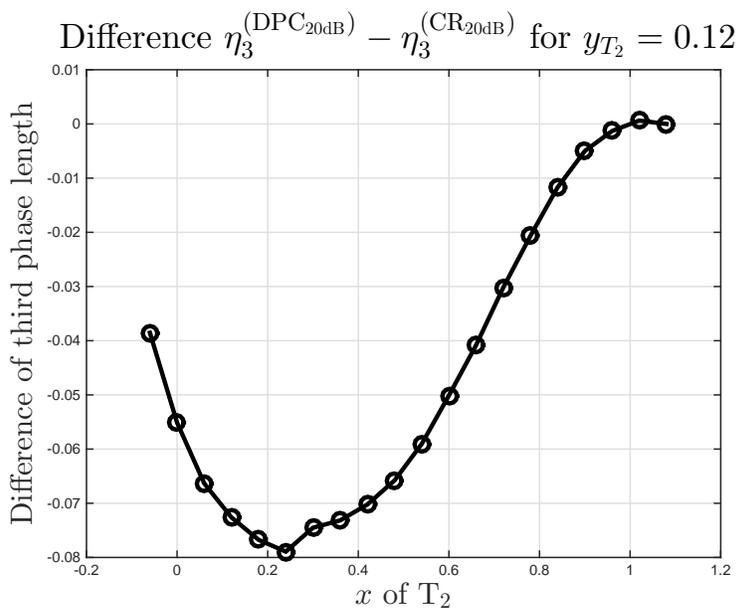}
\caption{Difference of optimal
$\eta_3^{(\text{DPC}_{20\text{dB}})}$ and
$\eta_3^{(\text{CR}_{20\text{dB}})}$ with $P_2=20$dB  as a function
of the position of T$_2$.}\label{fig:eta3_diff}
\end{figure}

\begin{figure}[t!]
\centering
\includegraphics[width=0.6\textwidth]{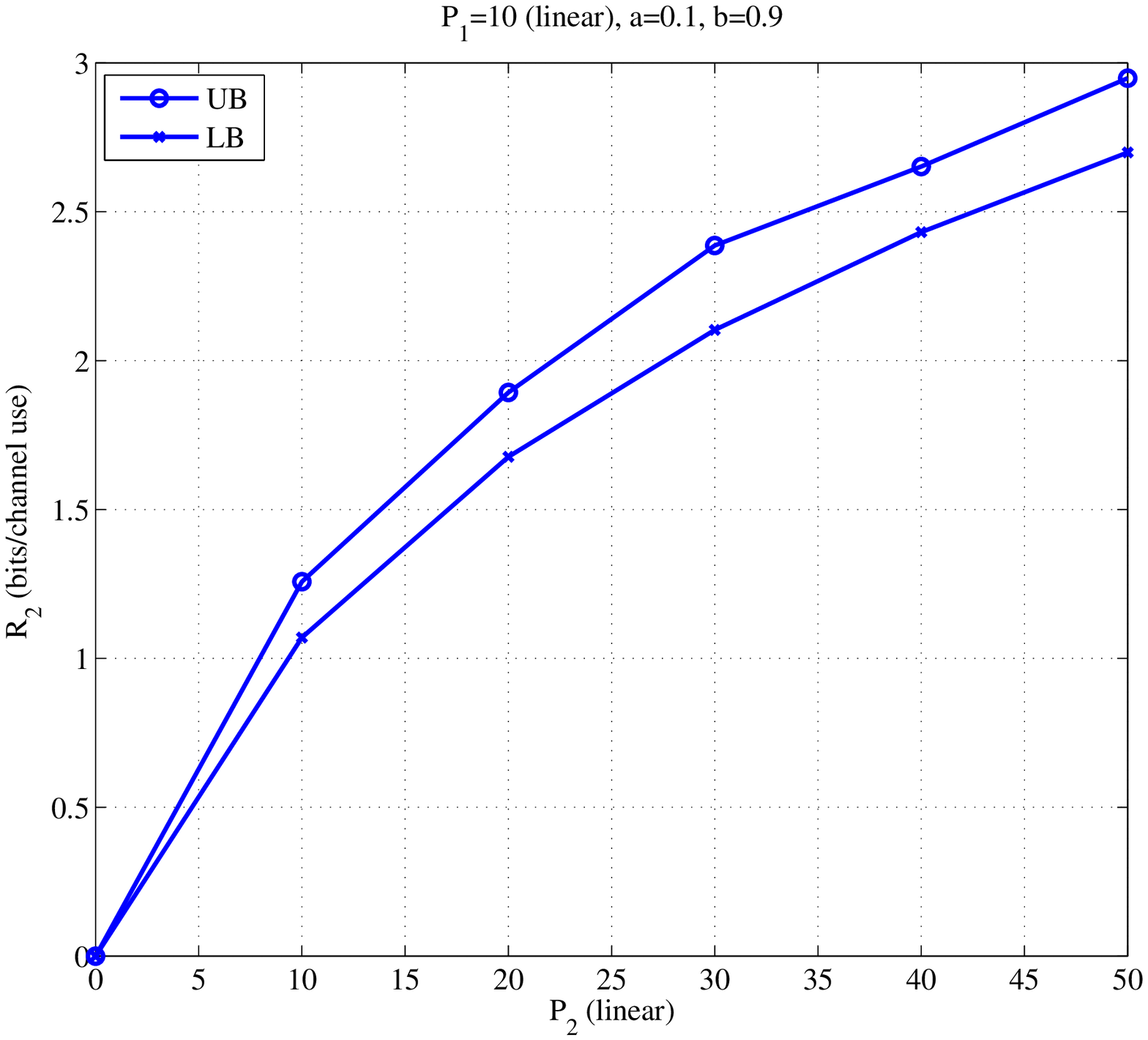}
\caption{The comparison of upper and lower bounds with $(a,b)=(0.1, 0.9)$.}\label{fig:comp_0109}
\end{figure}

We also investigate the interval of the third phase for clean
relaying depending on whether DPC is implemented or not, in Fig.
\ref{fig:eta3_DPC} and Fig. \ref{fig:eta3_diff}. Our results
highlight two important facts: First for both schemes, the 3-phase clean
relaying scheme will be used more often when T$_2$ gets closer to T$_1$.
This observation must be put in conjunction with the fact that
$\eta_1$ is getting smaller while T$_2$ gets closer to T$_1$, which
allows a longer period for clean relaying. Secondly, from the
difference of values of $\eta_3$ in Fig. \ref{fig:eta3_diff}, {one can
observe that the difference is always negative, which shows
that a longer "clean relaying" phase is needed when DPC is not
utilized. This is meaningful since clean relaying becomes more
crucial when the better transmission technique (DPC) is not used by
T$_2$. Thirdly, we observe that the difference admits a minimum,
i.e., (quantitatively, not rigorously) $\eta_3^{(CR_{20dB})}$ admits a maximum
for $x_{T_2}=0.2$, which coincides with the best location of a
relaying node. In other words, the time allocated to clean relaying
(without DPC) is larger at an optimized location of T$_2$, which is
in the interval $x\in[0.2, 0.3]$.} Note that this analysis can be corroborated by illustrating the power
consumption during the second and third phase depending on whether
DPC is used or not by T$_2$.
\end{enumerate}

\subsection{Comparison of upper and lower bounds}
{In this subsection we compare the derived inner and outer bounds. We consider
a channel with equivalent channel gains $(a,b)=(0.1,0.9)$ ($c_{12}=0.81$, $c_{21}=0.05$, $c_{TT}=10$, and $c_{11}=1$) with $P_1=10$ as an example.\footnote{Since here we fix $(a,b)$, changing $P_2$ will cause different $H_{22}$ due to a simple mapping between $(a,b)$ and $\{H_{12},H_{21},H_{22}\}$. Therefore, we do not show the set of $H_{22}$ to simplify the expression.} The \textit{smallest} ratio of the first phase
under the above setting is $\eta_1^*=0.3471$, which is the shortest interval that T$_2$ is able to successfully decode the primary message. Note that to fairly compare the upper and lower bounds, the upper bound is scaled by $1-\eta_1^*$ since the UB is derived for the channel that T$_2$ knows T$_1$'s message non-causally. Note also that we can not use an $\eta_1$ which is smaller than $\eta_1^*$ to be the scaling factor since which is feasible for T$_2$ to get the primary message. We observe that within $P_2=0\sim 50$ in linear scale, a gap between the upper and lower bounds is within 0.3 bits/channel use. That means the proposed achievable scheme is close to the capacity when the secondary transmitter/receiver is far/close enough to the primary receiver/transmitter, respectively. There might be two reasons for the existence of such gap. First, the upper bound is derived by neglecting the secure coexistence conditions, which might introduce some gap. Second, the constraints in the upper bound are much less than the ones in \cite{farsani_SCRC_ISIT}, \cite{farsani_CRC_arxiv}, \cite{farsani_unified_UB_IT} because part of the constraints are not adopted in our derivation. More specifically, some of the constraints in \cite{farsani_SCRC_ISIT}, \cite{farsani_CRC_arxiv}, \cite{farsani_unified_UB_IT} can be expressed by auxiliary random variables via simple change of variables due to the setting that the primary encoder is a deterministic one. But in our model we do not have the constraint of deterministic encoder at the primary transmitter. By degree of freedom (DoF) analysis with the assumption that $P_1$ is a finite constant derived in Appendix \ref{App_DoF_UB}, we can observe that the DoFs of the derived UB and LB is 1 and $\eta_2=1-\eta_1$, respectively, which in general do not match. However, when $|c_{TT}|\rightarrow\infty$, $\eta_1\rightarrow 0$ resulting in $\eta_2\rightarrow 1$, which matches the DoF of UB. \\
In summary, our numerical results highlight how the secondary user can securely coexist with the primary system while achieving positive rates for many geographical setups.

}

\section{Conclusions}\label{sec:conclusion}
In this paper we investigated cooperative secure communications in a
cognitive radio network where the secondary receiver is treated
as a potential eavesdropper with respect to the primary
transmission. Under secure coexistence conditions we derived
the secondary user's achievable rates and the related constraints to guarantee
the primary user's weak secrecy rate, when Gelfand-Pinsker coding is used.
Then we proposed a multi-phase transmission scheme to include the phases
for clean relaying with cooperative jamming and the
latency to successfully decode the primary message at the secondary transmitter. Additionally, an capacity upper bound of the secondary user is derived.
Numerical results illustrate that the
secondary transmitter can improve its performance by optimizing different parameters for the
transmission schemes according to the relative positions of the nodes. {We also show that the gap between our achievable scheme and a capacity upper bound can be within 0.3 bits/channel use in tested scenarios.} Thereby, a smart secondary transmitter is able
to adopt its relaying and cooperative jamming to guarantee primary
secrecy rates and transmit its own data at the same time from
relevant geometric positions.

\renewcommand{\thesection}{Appendix I}
\section{Proof of Theorem \ref{Th_conditions_single_binning_equals_double}} \label{App_conditions_single_binning_equals_double}
\begin{proof}
In the following we perform the reliability and equivocation rate
analysis.

\textit{Codebook generation:} We first generate $2^{n(R_{S1}+R_{S1}')}$
independent sequences $\bm v_1$ such that
$R_{S1}+R_{S1}'=I(V_1;Y_1)-2\epsilon,\,\forall\,\epsilon>0$ as the primary
user's codewords each with probability $P_{\bm V_1}(\bm
v_1)=\Pi_{i=1}^n P_{V_1}(v_{1i})$ and label them as
\begin{align}\label{EQ_codebook1}
\bm v_1(w_1,s_1),\,\, w_1\in\{1,\cdots,M_1\},\,\,
s_1\in\{1,\cdots,J_1\},
\end{align}
where $M_1=2^{nR_1}$ and $J_1=2^{nR_1'}$, $R_{S1}'\triangleq
I(V_1;Y_2)-2\epsilon$. We then generate $2^{n(R_2+R_2')}$
independent sequences $\bm v_2$ as the secondary user's codewords
each with probability $P_{\bm V_2}(\bm v_2)=\Pi_{i=1}^{n}
P_{V_2}(v_{2i})$ and label them as
\begin{align}
\bm v_2(w_2,s_2),\,\, w_2\in\{1,\cdots,M_2\},\,\,
s_2\in\{1,\cdots,J_2\},
\end{align}
where $M_2=2^{nR_2}$ and $J_2=2^{nR_2'}$, $R_2'\triangleq
I(V_1;V_2)+\epsilon$.

Let
\begin{align}
C_1\triangleq\{\bm v_1(w_1,s_1),\,\forall\,\,(w_1,s_1)\},\mbox{ and
}C_2\triangleq\{\bm v_2(w_2,s_2),\,\forall\,\,(w_2,s_2)\},\notag
\end{align}
be the codebooks of T$_1$ and T$_2$, respectively.\\
\textit{Encoding:} For the primary user to send the message $w_1$,
T$_1$ randomly chooses a codeword from bin $w_1$. For the secondary
user to send the message $w_2$, T$_2$ chooses the codeword in bin
$w_2$ such that
\begin{align}
(\bm v_1 (w_1,s_1),\bm v_2(w_2,s_2))\in
T_{\epsilon}^{(n)}(P_{V_1,V_2}),
\end{align}
where $T_{\epsilon}^{(n)}(P_{V_1,V_2})$ denotes the set of strongly
jointly typical sequences $\bm v_1 $ and $\bm v_2$ with respect to
the joint probability mass function of $V_1$ and $V_2$, i.e.,
$P_{V_1,V_2}$. The channel input $\bm x$ is selected by the
prefixing $P(x|v_1,v_2)$. Note that here we do not choose $\bm v_1 $
and $\bm v_2$ simultaneously as \cite{Liu_SISO}, but sequentially
just like the interference mitigated CR \cite{Jovicic_CR} since T$_1$'s coding scheme is kept intact here.\\

\textit{Decoding:} Decoder U$_1$ and U$_2$ choose $(w_1, s_1)$ and
$(w_2, s_2)$, respectively, so that
\begin{align}
(\bm v_1(w_1,s_1),\bm y_1')&\in T_{\epsilon}^{(n)}(P_{V_1,Y_1'}),\mbox{ and }(\bm v_2(w_2,s_2),{\bm y}'_{2}{ })\in T_{\epsilon}^{(n)}(P_{
V_2,Y_2'}),
\end{align}
if such $w_1$ and $w_2$ exist and if these are unique; otherwise, an error is
declared.
{ Note that this kind of jointly typical decoder is the same as the one used for the original U$_1$.}\\

\textit{Error analysis:} Without loss of generality, assume the messages $\bm v_1(1,1)$ and $\bm
v_2(1,1)$ are
transmitted. Three kinds of error should be considered: \\
$E_1$: T$_2$ cannot find any $\bm v_2$ which is jointly typical to
the side information $\bm v_1$. The probability of this error is denoted by $P_{E1}\triangleq Pr((\bm v_1,\bm v_2)\notin T_{\epsilon}^{(n)}),\,\forall \bm v_2$.\\
$E_2$: $\bm v_1(1,1)$ is not jointly typical to $\bm y_1'$ or $\bm
v_2(1,1)$ is not jointly typical to ${\bm y}'_{2}{ }$. The probability of this error is denoted by $P_{E2}\triangleq Pr((\bm v_1(1,1),\bm y_1')\notin T_{\epsilon}^{(n)}\cup (\bm v_2(1,1),\bm y_2')\notin T_{\epsilon}^{(n)})$.\\
$E_3$: $\bm v_1(w_1\neq 1,s_1)$ is jointly typical to $\bm y_1'$ or
$\bm v_2(w_2\neq 1,s_2)$ is jointly typical to ${\bm y}'_{2}{ }$.\\
For the event $E_1$, from the Covering Lemma \cite{Kim_book}, as
long as $R_2'>I(V_1;V_2)$, the probability $P_{E1}$ of this case
approaches to zero when $n$ is large enough. We then consider the
decoding error at the primary receiver. From joint typicality
\cite{Kim_book} we know that the probability $P_{E2}$ approaches to
zero when $n$ is large enough. For the event $E_3$, from
\cite{Kim_book} we know that
\begin{align}
P_{E3}=&P\{(\bm v_1(w_1\neq 1,s_1),\bm y_1')\in
T_{\epsilon}^{(n)}(P_{ V_1,Y_1'})\big|(\bm v_1 (1,1),\bm
v_2(1,1))\in
T_{\epsilon}^{(n)}(P_{V_1,V_2})\}
\leq  2^{-n[I( V_1;Y_1')-\epsilon]}.
\end{align}
Thus the total error probability can be bounded by the union bound as
\begin{align}
P_e&\leq P_{E1}\!+\!P_{E2}\!+\!\sum_{w_1\neq 1}\sum_{s_1}P_{E3}\leq P_{E1}\!+\!P_{E2}\!+\!M_1J_12^{-n[I( V_1;Y_1')-\epsilon]}\leq P_{E1}\!+\!P_{E2}+2^{-n[I(
V_1;Y_1')-R_{S1}-R_{S1}'-\epsilon]}.\label{EQ_error_analysis}
\end{align}
Thus to enforce the upper bound of $P_e$ to approach zero, it is
required that
\[
I( V_1; Y_1')>R_{S1}+R_{S1}'+\epsilon=I(V_1;Y_1)-\epsilon,
\]
which can be rearranged as $I( V_1;Y_1')\geq I(V_1;Y_1)$.

For the secondary user, we can follow the same steps to derive the
constraint $R_2<I(V_2;Y_2')-R_2'=I(V_2;Y_2')-I(V_2;V_1)-\epsilon$,
which
guarantees that the error probability approaches to zero. \\

\textit{Equivocation rate analysis:} 

The equivocation rate can be further rearranged as
\begin{align}
 H(W_1| \bm Y_2') \overset{(a)}\geq & H(  \bm V_1)\!-\!I( \bm V_1;
\bm V_2) \!-\! I(  \bm V_1;
\bm Y_2'|  \bm V_2) \!-\! H(  \bm V_1|  \bm V_2,\bm Y_2',W_1)\notag\\
 \overset{(b)}\geq &H(  \bm V_1) -n[I(V_1;V_2) +I( V_1;
Y_2' | V_2)+\epsilon] + H( \bm  V_1| \bm  V_2, \bm  Y_2',W_1)\notag\\
 \overset{(c)} =& H(  \bm V_1)-n[I( V_1; V_2,Y_2' )+\epsilon]+
H( \bm  V_1|  \bm V_2, \bm  Y_2',W_1)\notag\\
 \overset{(d)} =& n[R_{S1}+I(V_1;Y_2)-\epsilon]-n[I( V_1;
V_2,Y_2')+\epsilon]- H(  \bm V_1| \bm  V_2, \bm
Y_2',W_1),\label{EQ_EquivocationRateAnalysis}
\end{align}
where (a) is from \cite[(56)]{Liu_SISO},
{(b) is from \cite[Lemma 3]{Liu_SISO}, which is independent to the
channel model or coding scheme, but only relies on the basic
definitions of joint typicality, mutual information, and entropy.
Thus we can apply this lemma to our problem;} (c) uses the chain
rule of entropy again; (d) is due to $\bm V_1$ attaining
$2^{n\cdot(R_{S1}+I(V_1;Y_2)-\epsilon)}$ possible values with equal
probability from the construction of the code in
\eqref{EQ_codebook1}. Then we can rearrange
\eqref{EQ_EquivocationRateAnalysis} as
\begin{align}\label{EQ_EquivocationRateAnalysis2}
& R_{S1}-\frac{1}{n} H(W_1| \bm Y_2') \leq  I( V_1; V_2,Y_2' )
-I(V_1;Y_2)+2\epsilon+ \frac{1}{n}H( \bm V_1| \bm V_2, \bm
Y_2',W_1).
\end{align}
To achieve the weak secrecy, we need to make the RHS of
\eqref{EQ_EquivocationRateAnalysis2} approaches zero. We first
choose
\begin{equation}\label{EQ_final_Fano_valid_constraint}
 I( V_1; V_2,Y_2' )=I(V_1;Y_2).
\end{equation}
Then given $W_1=w_1$, we can apply Fano's inequality to the last
term on the RHS of \eqref{EQ_EquivocationRateAnalysis2} as
\begin{align}\label{EQ_fano}
H( \bm V_1| \bm V_2, \bm Y_2',W_1=w_1)\leq 1+\lambda\cdot n \cdot
I(V_1;V_2),
\end{align}
where $\lambda$ is the average probability of error of decoding the
index $s_1$ at U$_2$ given $ \bm V_2$ and $W_1$. We prove that
$\lambda$ approaches zero when $n$ is large enough in Appendix
II, then we have
\begin{align}\label{EQ_fano22}
\frac{1}{n}H( \bm V_1| \bm V_2, \bm Y_2',W_1=w_1)&\overset{(a)}\leq\frac{1}{n}H( \bm V_1| \bm Y_2',W_1=w_1)\overset{(b)}\leq
\frac{1}{n}(1+\lambda n I(V_1;Y_2'))\leq\epsilon_2,
\end{align}
where (a) is due to conditioning does not increase entropy and (b) is by Fano's inequality.
After averaging over $W_1$, we have
\begin{align}
&\frac{1}{n}H( \bm V_1| \bm V_2, \bm Y_2',W_1)=\frac{1}{n}\sum_{w_1}P(W_1=w_1)H( \bm V_1| \bm V_2, \bm Y_2',W_1=w_1)
\leq  \epsilon_2.\label{EQ_epsilon2}
\end{align}

Finally, after substituting \eqref{EQ_final_Fano_valid_constraint},
and \eqref{EQ_epsilon2} into
\eqref{EQ_EquivocationRateAnalysis2}, we have
\begin{align}\label{EQ_equivocation_rate}
R_{S1}-\frac{1}{n}h(W_1| \bm Y_2')\leq 2\epsilon+\epsilon_2.
\end{align}
This concludes the proof.
\end{proof}

\renewcommand{\thesection}{Appendix II}
\section{Proof of the constraint that $\lambda$ approaches zero} \label{App_Fano_for_single_binning}
\begin{proof}
In this appendix we use the following error analysis to prove
$\lambda<\epsilon$. Note that now we require that the primary user's
codeword, which is designed for single user secure transmission, is
jointly typical to the new received signal at $U_2$ when T$_2$
transmits. For a given typical sequence $\bm v_2$, let
$T_{\epsilon}^{(n)}(P_{ V_1, Y_2'| V_2})$ denote the set of jointly
typical sequences $\bm v_1$ and $\bm y_2'$ with respect to $P_{
V_1,{Y_2'}| V_2}$. Given $W_1=w_1$, U$_2$ chooses $s_1$ so that
\begin{align}
(\bm v_1(w_1,s_1),\bm y_2')\in T_{\epsilon}^{(n)}(P_{ V_1,
Y_2'| V_2}),
\end{align}
if such $s_1$ exists and is unique; otherwise an error is declared.
Define the event
\begin{align}\label{EQ_Def_conditional_JT}
E(s_1)=\{(\bm v_1(w_1,s_1),\bm y_2')\in T_{\epsilon}^{(n)}(P_{
V_1, Y_2'| V_2})\}.
\end{align}
Without loss of generality, we assume that $\bm v_1(w_1,1)$ is sent
and we define the event
\begin{align}
K_1=\{\bm v_1(w_1,1) \mbox{ was sent}\}.
\end{align}
Then we know with union bound that
\begin{align}\label{EQ_analysis_lambda}
\lambda(w_1)\leq P\{E^c(1)|K_1\}+\sum_{s_1\neq 1} P\{E(s_1)|K_1\},
\end{align}
where $E^c(1)$ denotes the event that $(\bm v_1(w_1,1),\bm
y_2')\notin T_{\epsilon}^{(n)}(P_{\bm V_1,\bm Y_2'|\bm V_2})$.

From joint typicality we know that
\begin{align}
P\{E^c(s_1=1)|K_1\}\leq \epsilon,\mbox{ and }P\{E(s_1\neq 1)|K_1\}\leq 2^{-n[I({V_1;Y_2'|V_2})-\epsilon]}.\label{EQ_prob_JT}
\end{align}

To find the upper bound of the second term on the RHS of
\eqref{EQ_analysis_lambda}, we first check the value of $H(\bm V_1|
\bm V_2, W_1)$, which represents the total number of cases that the decoding is wrong when deriving the total probability of error similar
to \eqref{EQ_error_analysis}, i.e.,
\begin{align}
& H( \bm V_1| \bm V_2, W_1)\notag\\
\overset{(a)}= &H( \bm V_1| \bm V_2)-I( \bm V_1;W_1| \bm V_2)\notag\\
\overset{(b)}=&H( \bm V_1) - I( \bm V_1; \bm V_2)-I( \bm V_1;W_1| \bm V_2)\notag\\
\overset{(c)}=& n(R_{S1}+R_{S1}'-\epsilon)- I( \bm V_1; \bm V_2)-I( \bm V_1;W_1| \bm V_2)\notag\\
\overset{(d)}\leq & n(R_{S1}+R_{S1}'-\epsilon)- n(I( V_1;
V_2)-\epsilon)-I(
\bm V_1;W_1| \bm V_2)\notag\\
\overset{(e)}= &n(R_{S1}+R_{S1}')- nI( V_1; V_2)-[H(W_1| \bm V_2)- H(W_1| \bm V_1, \bm V_2) ]\notag\\
\overset{(f)}= &n(R_{S1}+R_{S1}')- nI( V_1; V_2)-H(W_1| \bm V_2)
\label{EQ_number_of_uncertainty2}
%
\end{align}
where (a) and (b) are by the chain rule of entropy; (c) is by
definition of $ \bm V_1$ sequences, which are uniformly selected
with $2^{n(R_{S1}+R_{S1}'-\epsilon)}$ possibilities and
$R_{S1}'=I(V_1;Y_2)-2\epsilon$ as shown in the codebook construction;
(d) is due to the i.i.d assumption; (e) uses the chain rule of
entropy again; (f) uses the fact that when $\bm V_1$ is known, $W_1$
is also known, thus $H(W_1| \bm V_1, \bm V_2)=0$.

To proceed, we assume
\begin{align}\label{EQ_constraint_Pe_0}
I(V_1;Y_2)\geq I(V_1;V_2)-2\epsilon.
\end{align}
This assumption can be interpreted as follows. Given the primary
user's wiretap code, which is fixed, we enforce each bin has at
least one codeword jointly typical to $\bm v_2$, i.e.,
\begin{align}\label{EQ_constraint_equivocation}
2^{n[I(V_1;Y_2)-2\epsilon]}2^{-n[I(V_1;V_2)+\epsilon]} & =
2^{n[I(V_1;Y_2)-I(V_1;V_2)+3\epsilon]} \geq 1,
\end{align}
where $2^{n[I(V_1;Y_2)-2\epsilon]}$ is the number of codewords per
bin and $2^{-n[I(V_1;V_2)+\epsilon]}$ is the probability that the
codeword $\bm v_1$ is jointly typical to $\bm v_2$. This implies
$H(W_1|\bm V_2)=H(W_1)$, which is the same concept as the wiretap
coding. Thus \eqref{EQ_number_of_uncertainty2} becomes
\begin{align}
H( \bm V_1| \bm V_2, W_1) &=n[I(V_1;Y_2)-I( V_1; V_2)].\label{EQ_UB_conditional_entropy21}
\end{align}
Now we can bound the second term of error probability on the RHS of
\eqref{EQ_analysis_lambda} via \eqref{EQ_prob_JT} and
\eqref{EQ_UB_conditional_entropy21} as
\begin{align}\label{EQ_PER_check}
2^{n[I(V_1;Y_2)-I( V_1;
V_2)-2\epsilon]}2^{-n[I({V_1;Y_2'|V_2})-\epsilon]}
=2^{-n[I({V_1;Y_2',V_2})-I(V_1;Y_2)+\epsilon]}.
\end{align}
We can see that if \eqref{EQ_final_Fano_valid_constraint} is
fulfilled, \eqref{EQ_PER_check} approaches to zero when $n$ is large
enough. Now we compare the condition
\eqref{EQ_final_Fano_valid_constraint} and the assumption
\eqref{EQ_constraint_Pe_0}. We can express the LHS of
\eqref{EQ_final_Fano_valid_constraint} as
\begin{align}
I(V_1;V_2,Y_2')=I(V_1;V_2) + I(V_1;Y_2'|V_2) \overset{(a)}\geq I(V_1;V_2),
\end{align}
where (a) is due to $I(V_1;Y_2'|V_2)\geq 0$ \cite[Th.
2.34]{Yeung_IT_book}. Then combined with the RHS of
\eqref{EQ_final_Fano_valid_constraint}, we have $I(V_1;Y_2)\geq
I(V_1;V_2)$. Therefore, \eqref{EQ_final_Fano_valid_constraint}
 is sufficient to ensure \eqref{EQ_constraint_Pe_0} as $n\rightarrow\infty$, i.e., $\epsilon\rightarrow
 0$. Thus to make the probability of error \eqref{EQ_analysis_lambda} approach zero, we can find that \eqref{EQ_final_Fano_valid_constraint} is sufficient.

\end{proof}

\renewcommand{\thesection}{Appendix III}
\section{Proof of Proposition \ref{propDPC}} \label{App_DPC_AWGN}
\begin{proof}
The received signals at U$_1$ and U$_2$ for the three phases are
respectively given as
\begin{align}
Y_1'^{(1)} & = V_1^{(1)}+Z_1^{(1)},\notag\\
Y_1'^{(2)} & = X_1^{(2)}+c_{21}X_2^{(2)}+Z_1^{(2)}=c_{11}^{(2)}X_{1}^{(2)}+c_{21}(U_2^{(2)}+A_2^{(2)})+Z_1^{(2)},\notag\\
Y_1'^{(3)} & = c_{11}^{(3)}V_1^{(3)}+Z_1^{(3)},\notag\\
Y_2'^{(1)} & = c_{12}V_1^{(1)}+Z_2^{(1)},\notag\\
Y_2'^{(2)} & = c_{22}X_2^{(2)} + c_{12}X_1^{(2)}+Z_2^{(2)}= c_{12}^{(2)}X_1^{(2)} + c_{22}(U_2^{(2)}
+A_2^{(2)})+Z_2^{(2)},\notag\\
Y_2'^{(3)} & = c_{12}^{(3)}V_1^{(3)}+Z_2^{(3)},\notag
\end{align}
where $\phi_{21}$ is the phase of $c_{21}$, $X_2^{(2)} =U_2 +V_{1,r}
+A_2^{(2)} $, $V_{1,r} \triangleq
e^{-j\phi_{21}}\sqrt{(1-\rho_2)\gamma P_2^{(2)}/P_1}X_1^{(2)} $ is
T$_1$'s signal relayed by T$_2$, and $A_2^{(2)}$ is the jamming
signal in Phase 2. And $(V_1,V_2,U_2)$ have the following relation
\begin{align}
V_1 &= X_1,\,\,V_2 = U_2 +\alpha U_1,\,\,U_1 =
\left(c_{12}+c_{22}e^{-j\phi_{21}}\sqrt{\frac{(1-\rho_2)\gamma
P_2^{(2)}}{P_1}}\right)V_1=c_{12}^{(2)}V_1,
\end{align}
where $U_1$ is the side information for T$_2$ to precode, $X_1\sim
N(0,P_1)$, $\alpha=|c_{22}|^2 P_{U_2}/(1+ |c_{22}|^2 (P_{U_2} +
\rho_2 P_2^{(2)}))$ is the minimum mean square error (MMSE)
estimator of $U_2$ from the channel $Y=c_{22} (U_2+ A_2^{(2)})+N_2$,
where $A_2^{(2)} \sim \mathcal{N}\left(0,\rho_2 P_2^{(2)}\right)$ is
the cooperative jamming signal in the second phase, and
$P_{U_2}=(1-\rho_2)(1-\gamma) P_2^{(2)}$. By the construction of
DPC, $V_1$ is independent of $U_2$. In the third phase T$_2$ uses
the signaling
\begin{equation*}
X_2^{(3)}=
\sqrt{\frac{P_{2,1}^{(3)}}{P_1}}e^{-j\phi_{21}}X_1+A_2^{(3)},
\end{equation*}
where $A_2^{(3)} \sim \mathcal{N}\left(0,\rho_3 P_2^{(3)}\right)$ is
the cooperative jamming signal in the third phase. The equivalent noises at U$_1$ and U$_2$ in phases 1 and 3,
respectively, are $Z_1^{(1)}=Z_1$, $Z_2^{(1)}=Z_2$,
$Z_1^{(3)}=Z_1+c_{21}A_2^{(3)}$, $Z_2^{(3)}=Z_2+c_{22}A_2^{(3)}$. In the following
we calculate the terms in \eqref{EQ_old_new_main_channel2} and
\eqref{EQ_analysis_lambda2_constraint002} individually as
\begin{align}
I(V_1;Y_1^{(1)})&=\log(1+P_1^{(1)}),\label{EQ_I_1}\\
I(V_1;Y_2^{(1)})&=\log(1+|c_{12}|^2P_1^{(1)}),\label{EQ_I_2}\\
I(V_1;Y_1'^{(2)})&=\log\left(1+\frac{|c_{11}^{(2)}|^2P_1^{(2)}}{1+|c_{21}|^2(\rho_2 P_2^{(2)}+P_{U_2})}\right),\label{EQ_I_3}\\
I(V_1;V_2,Y_2'^{(2)})&=I(V_1;Y_2'|V_2)+I(V_1;V_2)\notag\\
&= I(V_1,V_2;Y_2')-\left(I(V_2;Y_2')-I(V_1;V_2)\right)\notag\\
&=\log\left(1+\frac{|c_{12}^{(2)}|^2P_1^{(2)}}{1+|c_{22}|^2(\rho_2 P_2^{(2)}+P_{U_2})}\right),\label{EQ_I_6}\\
I(V_1;V_2)&=h(U_2-\alpha V_1)-h(U_2-\alpha V_1 |V_1)=\log(1+\alpha^2 P_1/P_{U_2}),\label{EQ_I_8}\\
I(V_1;Y_1'^{(3)})&=\log\left(1+\frac{|c_{11}^{(3)}|^2P_1^{(3)}}{1+|c_{21}|^2\rho_3 P_2^{(3)}}\right),\label{EQ_I_9}\\
I(V_1;Y_2'^{(3)})&=\log\left(1+\frac{|c_{12}^{(3)}|^2P_1^{(3)}}{1+|c_{22}|^2\rho_3
P_2^{(3)}}\right).\label{EQ_I_10}
\end{align}

Therefore by substituting equations from
\eqref{EQ_I_1} to \eqref{EQ_I_10} into Proposition
\ref{Th_conditions_single_binning_equals_double}, we obtain
Proposition \ref{propDPC}.
\end{proof}

\renewcommand{\thesection}{Appendix IV}
\section{Proof of Theorem \ref{Thorem_3phase_gaussian_capacity}} \label{App_CT}
\begin{proof}
By the definitions \cite{Han_spectrum_book}
\begin{align}
p\mbox{-}\underset{n\rightarrow\infty}{\lim\inf}\frac{1}{n}i(\bm
X;\bm Y)&\triangleq
\sup\left\{\alpha:\lim_{n\rightarrow\infty}P\left(\frac{1}{n}i(\bm X;\bm Y)<\alpha\right)=0\right\},\notag\\
p\mbox{-}\underset{n\rightarrow\infty}{\lim\sup}\frac{1}{n}i(\bm
X;\bm Y)&\triangleq
\inf\left\{\alpha:\lim_{n\rightarrow\infty}P\left(\frac{1}{n}i(\bm
X;\bm Y)>\alpha\right)=0\right\},\notag
\end{align}
where $\bm X=[X_1,\cdots,X_n]$ and $\bm Y=[Y_1,\cdots,Y_n]$, the
secrecy capacity of a general wiretap channel can be restated as the
following from \cite[Corollary 1]{Bloch_strong_secrecy}
\begin{align}\label{EQ_Bloch_Cs}
C_s=\sup_{(\bm U,\bm X_1)\in \mathcal{P}_0}
\left(p\mbox{-}\underset{n\rightarrow\infty}{\lim\inf}\frac{1}{n}i(\bm
U;\bm
Y_1)-p\mbox{-}\underset{n\rightarrow\infty}{\lim\sup}\frac{1}{n}i(\bm
U;\bm Y_2)\right),
\end{align}
where $\mathcal{P}_0\triangleq\left\{\left\{\bm U\bm
X_1\right\}_{n\geq 1}:\forall\,n\in \mathds{N},\,\bm U\rightarrow
\bm X_1\rightarrow \bm Y_1\bm Y_2\right.$ forms a Markov chain and
$\frac{1}{n}c_n(\bm X_1)\leq P \mbox{ with}$\\
$\left.\mbox{probability }1\right\}$, $\left\{c_n\right\}_{n\geq 1}$
is a sequence of cost functions with $c_n: \mathcal{X}^n\mapsto
\mathds{R}^+$, and $i(\bm X;\bm Y)=\ln \frac{p(\bm X,\bm Y)}{p(\bm
X)p(\bm Y)}$ is the information density. Here we consider an
\textit{additive} cost constraint \cite[Sec.
3.6]{Han_spectrum_book}, i.e., $c_n(\bm x)=\sum_1^n c(x_i)$. Since
the whole channel is memoryless and for each phase the channel is
stationary, we can then rewrite the RHS of \eqref{EQ_Bloch_Cs} as
\begin{align}
C_s &\overset{(a)}=   \sup_{( \bm U ,\,\bm X_1 )\in \mathcal{P}_0}
\left( \underset{n\rightarrow\infty}{p\mbox{-}\lim\inf}\frac{1}{n}
\left\{i(\bm U^{(1)} ; \bm Y_1^{(1)})  +  i(\bm U^{(2)} ; \bm
Y_1^{(2)})  +  i(\bm U^{(3)};\bm
Y_1^{(3)})\right\}-\right.\notag\\
&\hspace{2.2cm}\left.
 \underset{n\rightarrow\infty}{p\mbox{-}\lim\sup}\frac{1}{n}\left\{i(\bm
U^{(1)} ; \bm Y_2^{(1)})  +  i(\bm U^{(2)} ; \bm Y_2^{(2)})
 +  i(\bm
U^{(3)} ; \bm Y_2^{(3)})  \right\}  \right)\notag\\
&\overset{(b)}=\sup_{(\bm U ,\,\bm X_1)\in \mathcal{P}_0}
\left(\underset{n\rightarrow\infty}{p\mbox{-}\lim\inf}\frac{1}{n}\left\{\overset{n_1}{\underset{j=1}{\Sigma}}
i(U^{(1)}_j;Y_{1j}^{(1)})+\overset{n_2}{\underset{j=1}{\Sigma}}i(U^{(2)}_j;Y_{1j}^{(2)})+\overset{n_3}{\underset{j=1}{\Sigma}}i(U^{(3)}_j; Y_{1j}^{(3)})\right\}-\right.\notag\\
&\hspace{2.2cm}\left.\underset{n\rightarrow\infty}{p\mbox{-}\lim\sup}\frac{1}{n}\left\{\overset{n_1}{\underset{j=1}{\Sigma}}
i(U^{(1)}_j;Y_{2j}^{(1)})+\overset{n_2}{\underset{j=1}{\Sigma}}
i(U^{(2)}_j;Y_{2j}^{(2)})+ \overset{n_3}{\underset{j=1}{\Sigma}} i(
U^{(3)}_j;
Y_{2j}^{(3)})\right\}\right)\notag\\
&\overset{(c)}=\sup_{(\bm U,\,\bm X_1)\in \mathcal{P}_0}
\left(\underset{n\rightarrow\infty}{p\mbox{-}\lim\inf}\left\{\frac{n_1}{n}\frac{1}{n_1}\overset{n_1}{\underset{j=1}{\Sigma}}
i(U^{(1)}_j;Y_{1j}^{(1)})+\frac{n_2}{n}\frac{1}{n_2}\overset{n_2}{\underset{j=1}{\Sigma}}i(U^{(2)}_j;Y_{1j}^{(2)})+\frac{n_3}{n}\frac{1}{n_3}\overset{n_3}{\underset{j=1}{\Sigma}}i(U^{(3)}_j; Y_{1j}^{(3)})\right\}-\right.\notag\\
&\hspace{2.2cm}\left.\underset{n\rightarrow\infty}{p\mbox{-}\lim\sup}\left\{\frac{n_1}{n}\frac{1}{n_1}\overset{n_1}{\underset{j=1}{\Sigma}}
i(U^{(1)}_j;Y_{2j}^{(1)})+\frac{n_2}{n}\frac{1}{n_2}\overset{n_2}{\underset{j=1}{\Sigma}}
i(U^{(2)}_j;Y_{2j}^{(2)})+
\frac{n_3}{n}\frac{1}{n_3}\overset{n_3}{\underset{j=1}{\Sigma}} i(
U^{(3)}_j;
Y_{2j}^{(3)})\right\}\right)\notag\\
\vspace{-0.2cm} &\overset{(d)}=\sup_{\{(U^{(k)},\, X_1^{(k)})\}\in
\mathcal{P}}
\left(\underset{n\rightarrow\infty}{p\mbox{-}\lim\inf}\left\{\frac{n_1}{n}
I(U^{(1)};Y_1^{(1)})+\frac{n_2}{n}I(U^{(2)};Y_1^{(2)})+\frac{n_3}{n}I(U^{(3)}; Y_1^{(3)})\right\}-\right.\notag\\
&\hspace{2.2cm}\left.\underset{n\rightarrow\infty}{p\mbox{-}\lim\sup}\left\{\frac{n_1}{n}
I(U^{(1)}; Y_2^{(1)})+\frac{n_2}{n} I(U^{(2)};Y_2^{(2)})+
\frac{n_3}{n} I( U^{(3)}; Y_2^{(3)})\right\}\right)\notag\\
\vspace{-0.2cm} &\overset{(e)}=\sup_{\{(U^{(k)},\, X_1^{(k)})\}\in
\mathcal{P}} \left(\left\{\eta_1
I(U^{(1)};Y_1^{(1)})+\eta_2I(U^{(2)};Y_1^{(2)})+\eta_3I(U^{(3)};
Y_1^{(3)})\right\}-\right.\notag\\
&\hspace{2.2cm}\left.\left\{\eta_1 I(U^{(1)}; Y_2^{(1)})+\eta_2
I(U^{(2)};Y_2^{(2)})+ \eta_3 I( U^{(3)};
Y_2^{(3)})\right\}\right)\notag
\end{align}
where in (a) we use the fact that there are three non-overlapped
phases and these phases are memoryless and independent; in (b) we
use the memoryless property $p_{\bm Y_1\bm Y_2|\bm
U}(y_1^n,y_2^n|u^n)=\Pi_{i=1}^np_{Y_1Y_2|X_1}(y_{1i},y_{2i}|x_{1i})\cdot$\\
$p_{X_1|U}(x_{1i}|u_i)$, and the fact that the distributions are
independent; in (c) we introduce $n_k/n_k$ for each phase for the
ease of the expression in average mutual information in the next
step; in (d) we apply law of large numbers: $\frac{1}{n_k}i(\bm
U^{(k)};\bm
Y_l^{(k)})=\frac{1}{n_k}\sum_{j=1}^{n_k}i(U^{(k)}_j;Y_{lj}^{(k)})\rightarrow
I(U^{(k)};Y_l^{(k)})\mbox{ a.s. as }n_k\rightarrow\infty $,
$k=1,\,2,\, 3$ and $l=1,\, 2$ and $\mathcal{P}$ is defined in
Theorem \ref{Thorem_3phase_gaussian_capacity}; in (e) we first
define $\eta_k\triangleq n_k/n$, $k=1\cdots 3$, which are fixed.
After substituting $\eta_k$, the RHS of (d) is independent of $n$
and we can remove the $p\mbox{-}\lim\inf$ and the
$p\mbox{-}\lim\sup$ operations. For the power constraint, we can
follow steps as in \cite[Theorem 3]{Bloch_strong_secrecy} with
discrete approximations to have the average power constraint. This
completes the proof.
\end{proof}

\renewcommand{\thesection}{Appendix V}
\section{Proof of Theorem \ref{Th_DMC_UB}} \label{App_DMC_UB}
{
\begin{proof}
Similar to \cite{Jovicic_CR}, we first relax the secure coexistence conditions to allow for joint code design for the primary and cognitive radio users. It results in an interference channel with degraded message sets with additional secrecy constraints. We start our proof from the result in \cite[Lemma 1]{farsani_CRC_arxiv} \cite[(14)]{farsani_unified_UB_IT}, which is an outer bound of a cognitive radio channel (or, an interference channel
with degraded message set) \textit{without} the coexistence condition but
with secrecy constraints on both primary and secondary messages. The
constraint set of the outer bound includes two parts: 1) a capacity
outer bound of a two-user peaceful cognitive radio channel from \cite{farsani_unified_UB_IT}; and 2)
equivocation rates constraints for each user.
\\

For the expressions of both parts, there are message variables and
time sharing variables in the mutual information expression, which may
hinder the simplification of the expression and also the further
specialization to AWGN cases. Especially, \cite{farsani_SCRC_ISIT}\cite{farsani_CRC_arxiv}\cite{farsani_unified_UB_IT} consider a set of joint PDFs where the channel input at
the primary transmitter is a deterministic function of primary message\footnote{The
reason that even a deterministic encoder is used but the secrecy
constraint can be fulfilled is that, they split the primary message
into two parts, one of them is transmitted by the cognitive
transmitter which uses a stochastic encoder. And this part of
message contributes a nonzero equivocation rate. }, by which they
can replace the message variable with the channel input variable
without affecting the mutual information and is convenient for
further manipulations, e.g., to derive the UB of AWGN case. The
simplified outer bound result is reported in \cite[Theorem 1]{farsani_SCRC_ISIT}\cite[Theorem 1]{farsani_CRC_arxiv}. However, there is no such constraint on the relation between the channel input and message at the primary transmitter in our model. Therefore, we
cannot
directly use the result from \cite[Theorem 1]{farsani_SCRC_ISIT}\cite[Theorem 1]{farsani_CRC_arxiv}. \\

In this proof, we re-derive the outer bound for our channel from
\cite[Lemma 1]{farsani_CRC_arxiv} as follows. Similar to \cite{farsani_CRC_arxiv}, we use the following auxiliary
random variables
\begin{align}\label{EQ_AUX}
U\triangleq (Z,\,M_1,\,Q), \,V\triangleq (Z,\,M_2,\,Q),
\,W\triangleq (Z,\,Q),
\end{align}
 where $Z\triangleq Z_Q$, $Q$ is a time sharing uniform random variable
  in the set $\{1,\cdots,n\}$ and $Z_t\triangleq (Y_1^{t-1},Y_{2,t+1}^n)$, $t=1,\cdots,n$.
   Note that constraints in \cite[Lemma 1]{farsani_CRC_arxiv}
 which are
conditioned by $\{M_1,\, Q\}$ or $\{M_2,\, Q\}$ are not considered
since which cannot be replaced by the auxiliary random variables in
(\ref{EQ_AUX}). In addition, because we do not have the deterministic relation between channel input and message at the primary transmitter, we are not able to replace $M_1$ by $X_1$ without changing the related mutual information in the upper bound constraints, which impedes the further derivation. With the
above criteria to select bounds/constraints from \cite[Lemma 1]{farsani_CRC_arxiv}, we
can have an intermediate upper bound as
\begin{equation}\label{EQ_UB_itermediate}
\mathcal{C}_o^{DMC1}=\bigcup_{(Q,M_1,M_2)\rightarrow (X_1,X_2) \rightarrow Z}\\\left\{\begin{array}{l}(R_1,R_2,R_{e_1})\in\mathds{R}^3_+:\notag\\
 R_{e_1}\leq R_1\notag\\
 R_1\leq I(M_1,Z;Y_1|Q)\notag\\
 R_2\leq I(M_2,Z;Y_2|Q)\notag\\
 R_1+R_2\leq I(M_1;Y_1|Z,M_2,Q)+I(M_2,Z;Y_2|Q)\notag\\
 R_{e_1}\leq I(M_1;Y_1|Z, M_2,Q)-I(M_1;Y_2|Z,M_2,Q)\end{array}\right\}.\\
\end{equation}

After plugging (\ref{EQ_AUX}) into $R_1$ and $R_2$, we have
\begin{align}
R_1&\leq I(M_1,Z;Y_1|Q)=I(M_1,Z,Q;Y_1|Q)\leq I(M_1,Z,Q;Y_1)=I(U;Y_1),\notag\\
R_2&\leq I(M_2,Z;Y_2|Q)=I(M_2,Z,Q;Y_2|Q)\leq I(M_2,Z,Q;Y_2)=I(V;Y_2),\notag
\end{align}
where the second inequality is due to the chain rule of mutual information with the fact that $I(Q;Y_1)$ and $I(Q;Y_2)$ are non-negative; the last equality is by definition of \eqref{EQ_AUX}.
The sum capacity constraint can be derived from \eqref{EQ_UB_itermediate} as follows
\begin{align}
R_1+R_2 &\leq I(M_1;Y_1|Z,M_2,Q)+I(M_2,Z;Y_2|Q)\notag\\
& \overset{(a)}{\leq} I(X_1;Y_1|Z,M_2,Q)+I(M_2,Z;Y_2|Q)\notag\\
& \overset{(b)}{=} I(X_1;Y_1|V)+I(M_2,Z;Y_2|Q)\notag\\
& \overset{(c)}{=} I(X_1;Y_1|V)+I(M_2,Z,Q;Y_2)-I(Q;Y_2)\notag\\
& \overset{(d)}{\leq} I(X_1;Y_1|V)+I(M_2,Z,Q;Y_2)\notag\\
& \overset{(e)}{=} I(X_1;Y_1|V)+I(V;Y_2),\notag
\end{align}
where (a) uses the Markov chain $M_1\rightarrow X_1\rightarrow Y_1$
and the data processing inequality; (b) uses \eqref{EQ_AUX}; (c)
uses the chain rule of mutual information; (d) uses the fact that
$I(Q;Y_2)\geq 0$; (e) again uses \eqref{EQ_AUX}.\\
The equivocation rate is derived as follows
\begin{align}
R_{e_1} &\leq I(M_1;Y_1|Z, M_2,Q)-I(M_1;Y_2|Z,M_2,Q)\notag\\
& = I(Z,M_1,Q;Y_1|Z, M_2,Q)-I(Z,M_1,Q;Y_2|Z,M_2,Q)\notag\\
& \overset{(a)}= I(U;Y_1|V)-I(U;Y_2|V)\notag\\
& \overset{(b)}= I(U,X_1;Y_1|V) - I(X_1;Y_1|U,V) - \left\{I(U,X_1;Y_2|V) - I(X_1;Y_2|U,V)\right\}\notag\\
& = I(U,X_1;Y_1|V) - I(U,X_1;Y_2|V)  - \left\{I(X_1;Y_1|U,V)- I(X_1;Y_2|U,V)\right\}\notag\\
& \overset{(c)}\leq I(U,X_1;Y_1|V) - I(U,X_1;Y_2|V)\notag\\
& \overset{(d)}= I(X_1;Y_1|V) - I(X_1;Y_2|V),
\end{align}
where (a) is by applying \eqref{EQ_AUX}; (b) is by the chain rule of
mutual information; (c) is by the degradedness condition, i.e.,
$Y_2$ is a degraded version of $Y_1$ with respect to $X_1$ such that secrecy can be guaranteed; (d) is due to the Markov chain
$U\rightarrow X_1 \rightarrow (Y_1,Y_2)$.
\end{proof}

\renewcommand{\thesection}{Appendix VI}
\section{Proof of Theorem \ref{Th_AWGN_UB}} \label{App_AWGN_UB}

\begin{proof}
In the following we derive the primary user's secrecy rate and the secondary user's rate for AWGN channels from \eqref{EQ_RS1_DMC} and \eqref{EQ_R2_selection}. We consider the following received
signals, which are transformed from the original one by the same way to generate \textit{the standard Gaussian IC} for complex cases \cite[Appendix A]{rini2}
\begin{align}
\tilde{Y}_1&=\tilde{X}_1+a\tilde{X}_2+Z_1,\,\,\tilde{Y}_2=b\tilde{X}_1+\tilde{X}_2+Z_2,\notag
\end{align}
where $\tilde{X}_1$ and $\tilde{X}_2$ are channel inputs of the primary and
cognitive transmitter with average power constraints, $\frac{1}{n}\sum_{i=1}^n
|\tilde{X}_{1,i}|^2\leq \tilde{P}_1=|H_{11}|^2P_{1}$ and $\frac{1}{n}\sum_{i=1}^n |\tilde{X}_{2,i}|^2\leq \tilde{P}_2=|H_{22}|^2P_{2}$,
respectively, $a\triangleq \frac{h_{21}}{h_{22}}e^{j(-\angle h_{11}+\angle h_{21})}$, $b\triangleq \frac{|h_{12}|}{|h_{11}|}$ and $Z_1$ and $Z_2$ are circularly symmetric independent complex AWGN variables with zero
mean and unit variances, respectively. The UB of $R_1$ can be first derived by
\begin{align}
R_1&\leq I(U;\tilde{Y}_1)=h(\tilde{Y}_1)-h(\tilde{Y}_1|U)\notag\\
&\overset{(a)}=h(\tilde{Y}_1)- \log\left({1+\alpha(\tilde{P}_1+|a|^2\tilde{P}_2+2\Re\{a\rho\}\sqrt{\tilde{P}_1\tilde{P}_2})}\right)\notag\\
&\overset{(b)}\leq \log\left(\frac{1+\tilde{P}_1+|a|^2\tilde{P}_2+2\Re\{a\rho\}\sqrt{\tilde{P}_1\tilde{P}_2}}{1+\alpha(\tilde{P}_1+|a|^2\tilde{P}_2+2\Re\{a\rho\}\sqrt{\tilde{P}_1\tilde{P}_2})}\right),\label{EQ_R1_AWGN}
\end{align}
where (a) is by considering upper and
lower bounds of $h(\tilde{Y}_1|U)$ as follows
\begin{align}
h(\tilde{Y}_1|U)&\leq h(\tilde{Y}_1)\leq  \log2\pi e\left(1+\tilde{P}_1+|a|^2\tilde{P}_2+2\Re\{a\rho\}\sqrt{\tilde{P}_1\tilde{P}_2}\right),\notag\\
h(\tilde{Y}_1|U)&\geq h(\tilde{Y}_1|U,\tilde{X}_1,\tilde{X}_2) =  \log2\pi e.\notag
\end{align}
From the above it is obvious that there exists $0\leq \alpha\leq 1$ such that the following equality is valid
\begin{align}
h(\tilde{Y}_1|U)=  \log2\pi e\left(1+\alpha(\tilde{P}_1+|a|^2\tilde{P}_2+2\Re\{a\rho\}\sqrt{\tilde{P}_1\tilde{P}_2})\right).\label{EQ_h_Y1_cond_U}
\end{align}
(b) comes from the fact that Gaussian distribution maximizes entropy given a second moment.

To derive a tight enough upper bound for $R_{e1}$, we use the same trick as in \eqref{EQ_h_Y1_cond_U}. The UB of the first term of $R_{e1}$, i.e., $I(\tilde{X}_1;\tilde{Y}_1|V)$ can be derived by
\begin{align}
I(\tilde{X}_1;\tilde{Y}_1|V)&= \log\left(1+\gamma(\tilde{P}_1+|a|^2\tilde{P}_2+2\Re\{a\rho\}\sqrt{\tilde{P}_1\tilde{P}_2})\right)-\log(1+\eta |a|^2 \tilde{P}_2), \,\,0\leq\gamma,\,\eta\leq 1.\label{EQ_I_X1_Y1_given_V}
\end{align}

Similarly, we can derive the following expression for $I(\tilde{X}_1;\tilde{Y}_2|V)$ as
\begin{align}
I(\tilde{X}_1;\tilde{Y}_2|V)= \log\left(1+\beta(|b|^2\tilde{P}_1+\tilde{P}_2+2\Re\{b\rho\}\sqrt{\tilde{P}_1\tilde{P}_2})\right)-\log(1+\delta  \tilde{P}_1), \,\,0\leq\beta,\,\delta\leq 1.\label{EQ_I_X1_Y2_given_V}
\end{align}
After subtracting \eqref{EQ_I_X1_Y2_given_V} from \eqref{EQ_I_X1_Y1_given_V}, we have
\begin{align}
R_{e1}\leq  \Bigg(\log\left(\frac{1+\gamma(\tilde{P}_1+|a|^2\tilde{P}_2+2\Re\{a\rho\}\sqrt{\tilde{P}_1\tilde{P}_2})}
{1+\beta(|b|^2\tilde{P}_1+\tilde{P}_2+2\Re\{b\rho\}\sqrt{\tilde{P}_1\tilde{P}_2})}\right)-\log\left(\frac{1+\eta |a|^2 \tilde{P}_2}{1+\delta  \tilde{P}_1}\right)\Bigg)^+.\label{EQ_Re1_AWGN}
\end{align}

From \eqref{EQ_R1_AWGN} and \eqref{EQ_Re1_AWGN} we can derive $R_{s1}$ as
\begin{align}
R_{s1}\leq &\min\{R_1,\,R_{e1}\}\notag\\
=&\min\Big\{ \log\left(\frac{1+\tilde{P}_1+|a|^2\tilde{P}_2+2\Re\{a\rho\}\sqrt{\tilde{P}_1\tilde{P}_2}}{1+\alpha(\tilde{P}_1+|a|^2\tilde{P}_2+2\Re\{a\rho\}\sqrt{\tilde{P}_1\tilde{P}_2})}\right),\notag\\
&\hspace{0.9cm} \Bigg(\log\left(\frac{1+\gamma(\tilde{P}_1+|a|^2\tilde{P}_2+2\Re\{a\rho\}\sqrt{\tilde{P}_1\tilde{P}_2}}
{1+\beta(|b|^2\tilde{P}_1+\tilde{P}_2+2\Re\{b\rho\}\sqrt{\tilde{P}_1\tilde{P}_2})}\right)-\log\left(\frac{1+\eta |a|^2 \tilde{P}_2}{1+\delta  \tilde{P}_1}\right)\Bigg)^+\Big\}.
\end{align}

Two upper bounds of $R_2$ can be derived: similar to \eqref{EQ_R1_AWGN} we can derive
\begin{align}
R_2\leq &I(V;\tilde{Y}_2)\leq
 \log\left(\frac{1+|b|^2\tilde{P}_1+\tilde{P}_2+2\Re\{b\rho\}\sqrt{\tilde{P}_1\tilde{P}_2}}{1+\beta(|b|^2\tilde{P}_1+\tilde{P}_2+2\Re\{b\rho\}\sqrt{\tilde{P}_1\tilde{P}_2})}\right),\label{EQ_R2_UB1}
\end{align}
and we can derive from the sum-rate constraint
\begin{align}
R_2&\overset{(d)}\leq I(\tilde{X}_1;\tilde{Y}_1|V)+I(V;\tilde{Y}_2)-R_{S1, target}\notag\\
&\leq  \log\!\!\left(\frac{1+\gamma(\tilde{P}_1+|a|^2\tilde{P}_2+2\Re\{a\rho\}\sqrt{\tilde{P}_1\tilde{P}_2})}{1+\eta |a|^2 \tilde{P}_2}\!\!\right)\!\!+\log\!\!\left(\!\!\frac{1+|b|^2\tilde{P}_1+\tilde{P}_2+2\Re\{b\rho\}\sqrt{\tilde{P}_1\tilde{P}_2}}{1+\beta(|b|^2\tilde{P}_1+\tilde{P}_2+2\Re\{b\rho\}\sqrt{\tilde{P}_1\tilde{P}_2})}\!\!\right)\!\!-\log\!\!\left(\!\!\frac{1+\tilde{P}_1}{1+|b|^2\tilde{P}_2}\right),\label{EQ_R2_UB2}
\end{align}
where in (d) we exploit the Fourier-Motzkin elimination with the fact that $R_1\geq R_{S1,target}$ due to the secure coexistence condition (i). Comparing \eqref{EQ_R2_UB1} and \eqref{EQ_R2_UB2} concludes the proof.
\end{proof}
}

\renewcommand{\thesection}{Appendix VII}
\section{Proof of DoF of UB and LB} \label{App_DoF_UB}
we identify this fact by deriving the degree of freedom (DoF) of lower and upper bounds to investigate the gap in between. We consider the case in which $P_1$ is fixed with a finite value and $P_2$ approaches infinity. In this case, T$_2$ can use $\eta_3\rightarrow 0$ to satisfy the secure coexistence conditions by using a large but finite $P_2^{(3)}$. The derivation is as follows. For UB, we can derive the DoF of $R_2$, namely, $DoF_{2}^{UB}$, as:
\begin{align}
DoF_{2}^{UB}&\triangleq\lim_{P_2\rightarrow \infty} \frac{R_{2}^{UB}}{\log P_2}\notag\\
&\overset{(a)}\leq \lim_{P_2\rightarrow \infty}\,\,\,\,\, \Bigg\{\underbrace{\log\left(1+|b|^2\tilde{P}_1+\tilde{P}_2+2\Re\{b\rho\}\sqrt{\tilde{P}_1\tilde{P}_2}\right)}_{(a1)}\notag\\
&\hspace{1.2cm}-\Bigg(\underbrace{\log\left(\frac{1+\tilde{P}_1}{1+|b|^2\tilde{P}_2}\right)}_{(a2)}-\underbrace{\log\left(\frac{1+\gamma(\tilde{P}_1+|a|^2\tilde{P}_2+2\Re\{a\rho\}\sqrt{\tilde{P}_1\tilde{P}_2})}{1+\eta |a|^2 \tilde{P}_2}\right)}_{(a3)}\Bigg)^+\Bigg\}\Bigg/\log P_2\label{EQ_DoF_R2}\\
&=1\notag,
\end{align}
where in (a), $\tilde{P}_2=|H_{22}|^2P_2$ as defined in Theorem 4. In addition, we set $\beta=0$ in Theorem 4 to get (a1). This is because we need to maximize $R_2$ and if we do not choose $\beta=0$, the first term of the UB of $R_2$ in Theorem 4 will be zero when $P_2\rightarrow\infty$ since whose numerator and denominator both have the term $\tilde{P}_2$. In addition,
\[
\lim_{P_2\rightarrow\infty} \frac{((a2)-(a3))^+}{\log P_2}=0,
\]
no matter $\gamma$ and/or $\eta$ are zeros or not.
Then it is clear that $DoF_{2}^{UB}=1$.

From (15), the DoF of the LB of $R_2$, namely, $DoF_{2}^{LB}$, can also be derived as:
\begin{align}
DoF_{2}^{LB}&\triangleq\lim_{P_2\rightarrow \infty} \frac{R_{2}^{LB}}{\log P_2}\notag\\
&\leq \lim_{P_2\rightarrow \infty} \frac{\eta_2 \left(\log\left(1+|c_{22}|^2(1-\gamma+\rho_2\gamma)P_2^{(2)}\right)-\log\left(1+|c_{22}|^2\rho_2P_2^{(2)}\right)\right)}{\log P_2}\label{EQ_DoF_R2_LB}.
\end{align}
Recall that $\rho_2$ is the fraction of the power for jamming. We can observe that when $\rho_2=0$ and the constraints (16) and (17) in the manuscript are fulfilled, $DoF_{2}^{LB}\leq \eta_2$, where $\eta_2$ is non-zero. This case can happen if T$_2$ helps the primary user's secure transmission more in the third phase, e.g., by introducing more jamming power, instead of using part of $P_2^{(2)}$ to jam U$_2$ in the second phase. From numerical results under high $P_2$, e.g., 40 dB, we can find that $\rho_2=0$ for all positions we considered as in Fig. 2. For normal value of $c_{TT}$, we have $\eta_1>0$. Because $\eta_2=1-\eta_1-\eta_2\rightarrow 1-\eta_1$, where the last step is due to the assumption of finite $P_1$ which results in $\eta_3\rightarrow 0$ as explained in the beginning of the response to Comment 5, we know that the DoFs of the LB and UB do not match. However, if $|c_{TT}|\rightarrow \infty$, we will have $\eta_1\rightarrow 0$, then $\eta_2\rightarrow 1$, which coincides with the UB.

\bibliographystyle{IEEEtran}
\renewcommand{\baselinestretch}{1.8}
\bibliography{IEEEabrv,journal,thesis_ref}

\end{document}